\documentclass[amsmath,amssymb,nofootinbib,notitlepage,superscriptaddress,twocolumn,floatfix]{revtex4-2}

\usepackage{amsmath,graphicx,mathtools,setspace}
\usepackage[margin=1in,letterpaper]{geometry}
\usepackage[colorlinks=true]{hyperref}

\usepackage{booktabs}
\usepackage{array}
\usepackage{xcolor}
\usepackage{rotating,array}

\definecolor{GFblue}{RGB}{198,218,255}

\usepackage[bb=dsserif]{mathalpha}
\usepackage{bm}

\newtagform{brackets}{[}{]}
\usepackage{amsfonts}
\usepackage{color}
\usepackage{graphicx}
\usepackage{amsmath}
\usepackage{amsthm}
\usepackage{mathrsfs}
\usepackage{bm}
\usepackage{comment}
\usepackage{braket}
\usepackage{ulem}
\usepackage[export]{adjustbox}

\newcommand{\pd}{\partial}

\usepackage[T1]{fontenc}

\begin{document}

\title{Amplification of metric perturbations by extremal horizons}
\author{Samuel E. Gralla}
\author{Sepehr Salamat}
\affiliation{Department of Physics, University of Arizona, Tucson, AZ 85721, USA}

\begin{abstract}
Under normal circumstances, an observer falling into a black hole feels nothing special upon crossing the horizon.  In this paper we find an intriguing exception: For most of the parameter range of the extremal Kerr-Newman (KN) spacetime, horizon co-rotating perturbations are enhanced near the horizon, such that an infalling observer experiences an anomalously large tidal deformation as they enter the black hole.  Such perturbations arise when there is a persistent source outside the black hole that is either co-rotating itself or has discrete Fourier support at the associated frequencies $\omega=m\Omega_H$ (where $m$ is the azimuthal number and $\Omega_H$ is the horizon frequency).  The enhancement is formally infinite at precise co-rotation, and we work with a nearly co-rotating mode to keep perturbation theory under control.  The underlying physics is the emergence of discrete self-similarity in the extremal limit, with complex scaling weights of the form $-1/2\pm i\alpha$ for real $\alpha$.  It is analogous to the black hole Meissner effect, except that the near-horizon field is enhanced rather than screened.  We numerically calculate the scaling exponents for coupled gravitoelectromagnetic (GEM) perturbations of extremal KN black holes and show that the complex exponents arise in the parameter range $Q < Q_*$ with $Q_*\approx.93M$.  These exponents also predict the decay and growth rates (Aretakis effect) of generic GEM perturbations of the KN spacetime, both on and off the horizon.  The enhancement of co-rotating perturbations can be viewed as a driven Aretakis instability.
\end{abstract}

\maketitle

\section{Introduction}

A famous prediction of general relativity is that an observer entering a macroscopic black hole is oblivious to his impending doom, feeling nothing special at the moment of horizon crossing.  An exception to this rule is the Aretakis instability of extremal black holes \cite{Aretakis:2010gd, Aretakis:2011ha, Aretakis:2011hc, Aretakis:2011gz, Aretakis:2013dpa, Aretakis:2012bm, Lucietti:2012xr, Murata:2013daa, Aretakis:2012ei, Casals:2016mel, Gralla:2016sxp, Zimmerman:2016qtn, Gralla:2017lto, Gralla:2018xzo, Gralla:2018xoz, Hadar:2018izi, Angelopoulos:2019gjn, Gralla:2019isj, Angelopoulos:2018uwb, Apetroaie:2022rew, Gajic:2023uwh, Pourhassan:2025tqm, Chen:2025sim, Porfyriadis:2025pov, Angelopoulos:2024yev, Gelles:2026bix}, which implies that observers momentarily experience large curvature as they cross the horizon; however, the tidal deformation remains small.  In this paper we will discuss a related phenomenon in which infalling observers do experience noticeable tidal deformation.  We show that nearly co-rotating perturbations of extremal black holes are effectively amplified near the horizon, such that infalling observers experience a sudden increase in tidal deformation as they cross.  At precise co-rotation the tidal deformation is formally infinite, heralding a breakdown of perturbation theory in a regime where it would naively be expected to be valid.  

The construction is inspired by recent work  showing that the smoothness of stationary, axisymmetric extremal horizons in general relativity is in some sense accidental \cite{Horowitz:2022mly,Horowitz:2023xyl,Horowitz:2024dch,Horowitz:2024kcx,Horowitz:2026axions}.  While that body of work found singularities by incorporating effective field theory corrections or axion fields, we will find analogous behavior within ordinary Einstein-Maxwell theory by going beyond stationarity and axisymmetry.  The key point is that horizon co-rotating perturbations (sometimes called ``synchronous'' \cite{Richartz:2017qep}) enjoy near-horizon self-similarity in the same way as stationary, axisymmetric perturbations, while allowing a much richer behavior, including fractional and complex scaling exponents.  

The potential ``danger'' of co-rotating perturbations can be seen immediately from the Teukolsky equation \cite{teukolsky1973perturbations}.  After enforcing the co-rotating condition $\omega=m\Omega_H$ (with $\Omega_H$ the horizon frequency and $m$ the azimuthal mode number), the usual wave behavior at the extremal horizon $r_H=M$ disappears and is instead replaced by power laws in horizon deviation $x=(r-M)/M$ (see \cite{Teukolsky:1974yv,Richartz:2017qep} or Eq.~\eqref{psicorotHH} below),
\begin{align}\label{Danger}
    {}_s\hat{R}_{\ell m (m\Omega_H)}\sim ax^{h-1+s+im} + b x^{-h+s+im}.
\end{align}
Here ${}_s\hat{R}_{\ell m(\omega)}$ is the radial dependence of a term in the mode sum for a Weyl scalar $\psi_s$ done in horizon-regular coordinates and a horizon-regular tetrad (precise definitions given in the main body), and $h$ is a real or complex number, the ``conformal weight'' of the mode.  

In the presence of power-law end behavior, one normally eliminates a  singular power by imposing regularity.  However, the Kerr spacetime has ``principal modes'' where $h=1/2+i\delta$ for real $\delta$.  In this case both terms scale as $x^{-1/2+s}$; they are indistinguishable based on regularity, and both are singular for $s \leq 0$.  Choosing $s=-2$ corresponds to a component of spacetime curvature seen by an infalling observer, meaning that an observer would see curvature diverging as $x^{-5/2}$.  The net tidal deformation $\Delta \xi$ scales like two integrals, so $\Delta \xi \sim x^{-1/2}\to \infty$.  There are two logical possibilities: either these modes are somehow impossible to excite, or linearized perturbations can produce large tidal deformation.

\begin{table*}[htb]
\centering
\begin{tabular}{lccc}
\hline\hline
Perturbation Type & Self-Similar? & Weight & Physical Implication \\
\hline
Stationary, axisymmetric & Yes & Positive integer & Meissner-type screening \\
Stationary, nonaxisymmetric & No & N/A & No distinctive extremal behavior \\
Co-rotating (synchronous) & Yes & $\exists$ Negative real part & Singular horizon \\
Freely evolving (smooth initial data) & Yes & Positive real part & Aretakis behavior \\
\hline\hline
\end{tabular}
\caption{Summary of perturbation types, their near-horizon scaling self-similarity, associated weights, and resulting physical implications.}\label{tab:summary}
\end{table*}

To elucidate the situation, we consider \textit{nearly} co-rotating modes, where the standard ingoing-wave conditions can be imposed.  In this case the power laws \eqref{Danger} arise in the regime $k \ll x \ll 1$, where $k=2(\omega-m\Omega_H)/\Omega_H$ is the deviation from precise co-rotation (with a conventional factor of $2$).  We show that the modes of the retarded Green function become arbitrarily large as $k \to 0$, with the same scaling $k^{-1/2}$, confirming the presence of fields that produce large tidal deformation.  The analysis is done analytically for the Kerr spacetime and numerically for gravitoelectromagnetic (GEM) perturbations of the KN spacetime (mass $M>0$ and charge $Q\geq0$).  Since the region $k \ll x \ll 1$ of large field approaches the horizon as $k \to 0$, we say as a shorthand that the horizon is singular. 

These singular modes can be excited physically using a co-rotating source, or at least a source with discrete Fourier support at the co-rotating modes $\omega=m\Omega_H$. In the parameter range $Q/M>\sqrt{3}/2$ there are timelike co-rotating equatorial orbits, so a single orbiting particle suffices.  For lower black hole charge, one can instead consider a non-circular orbit tuned to have a non-zero co-rotating component.  We expect that such a construction is always possible, and provide a definite example in the Kerr spacetime.  We also show how to build a precisely co-rotating source from a swarm of such particles.  For purely electromagnetic perturbations of Kerr, the source need not execute a geodesic motion, so one may choose a point charge executing a small circular motion about the symmetry axis.

These results can be understood geometrically in terms of the invariant \textit{weight} of a tensor field under near-horizon scaling \cite{gralla2016near,Gralla:2017lto}.  In Ref.~\cite{gralla2016near} this concept was used to understand the black hole Meissner effect \cite{bivcak1985magnetic,king1975black,chamblin1998superconducting,karas1991interpretation,karas2000magnetic,bini2008charged,semerak2002expulsion,penna2014black,penna2014blackentanglement,bivcak2015near,kofrovn2016separability,gralla2016near,gurlebeck2017meissner,kunz2017magnetized,gurlebeck2018meissner,giribet2023zooming}, showing that the \textit{positive integer} weight of stationary, axisymmetric perturbations gives rise to the characteristic screening of EM fields.\footnote{Refs.~\cite{gralla2016near,Gralla:2017lto} use an opposite sign convention for the definition of weight; see below \eqref{wwtilde} for details.}  Later, the notion of weight was applied to the Aretakis instability, where the weight was identified with the late-time power-law decay or growth, with the \textit{positive real part} indicating that scalar invariants decay \cite{Gralla:2017lto}.  This paper shows that co-rotating modes have self-similar weight with \textit{negative real part} ($w=-1/2+i \alpha$ for $\alpha\in\mathbb{R}$), indicating singular near-horizon behavior.  We utilize the interrelationships between these phenomena (Tab.~\ref{tab:summary})  to confirm the presence of a gravitational Meissner effect and make predictions for the precise decay and growth rate of GEM perturbations of the KN spacetime.

Our analysis shows that there exist linearized solutions of the sourced Einstein-Maxwell equations displaying near-horizon amplification, with observable consequences for infalling observers.  However, the construction requires sources that contain exactly co-rotating harmonics (e.g., a particle on a particular bound orbit for all time), which neglects radiation reaction and can only be regarded as a local-in-time approximation to an evolving spacetime.  Furthermore, any evolution toward the co-rotating state will presumably evolve the black hole away from an initially extremal state.  Further work is required to estimate the degree of amplification that is realizable with smooth evolution from initial data in the full nonlinear theory.

This paper is organized as follows.  In Sec.~\ref{sec:NP} we review the perturbation framework and introduce notation.  In Sec.~\ref{sec:Kerr} we consider co-rotating perturbations of the Kerr spacetime, and in Sec.~\ref{sec:KN} we consider co-rotating perturbations of the KN spacetime.  We then review the notion of tensor weight in Sec.~\ref{sec:weight} and use it to connect with the Meissner and Aretakis effects in Secs.~\ref{sec:Meissner} and \ref{sec:Aretakis}, respectively.  Our metric signature is $(-+++)$ and we use units with $G=c=4\pi \epsilon_0=1$. 

\section{Newman-Penrose perturbation framework}\label{sec:NP}

The Kerr-Newman (KN) metric \cite{kerr1963gravitational,newman1965metric} is reviewed in  \cite{adamo2014kerr}.  We consider only the extremal case,
 \begin{align}
     a = \sqrt{M^2-Q^2},
 \end{align}
 where $M>0$ is the mass, $Q\geq0$ is the charge, and $a\geq0$ is the spin parameter (angular momentum $aM$).  The horizon radius and angular frequency are
 \begin{align}
     r_H=M, \qquad \Omega_H=\frac{a}{M^2+a^2}.
 \end{align}
 We will also find it useful to introduce
 \begin{align}
     \Delta = (r-M)^2, \qquad \rho = r+ i a \cos \theta.
 \end{align}

 We will be using the Newman-Penrose (NP) approach to black hole perturbations \cite{newman1962approach,teukolsky1973perturbations} throughout this paper.  We form the standard NP quantities suitable for perturbations of the KN spacetime \cite{chandrasekhar1998mathematical,dias2015linear,mark2015quasinormal}, 
\begin{align}
        \psi_{-1}& =  \frac{3(\rho^*)^2(M\rho - Q^2)}{\sqrt{2}\,Q\rho}  \delta \phi_{2} \\ & = \frac{3(\rho^*)^2(M\rho - Q^2)}{\sqrt{2}\,Q\rho}  \delta F_{\mu \nu}  m^*{}^\mu n^\nu  \label{psim1} \\ 
    \psi_{-2} & = \rho^*{}^4 \Psi_4 = \rho^*{}^4 \delta C_{\mu \nu \rho \sigma} n^\mu m^*{}^\nu n^\rho m^*{}^\sigma,\label{psim2}
\end{align}
presented here in the gauge with $\Psi_3=0$.  Here $\delta F_{\mu \nu}$ and $\delta C_{\mu \nu \rho \sigma}$ are the perturbations to the field strength and Weyl tensor, respectively.

For calculations we use the Kinnersley tetrad in Boyer-Lindquist (BL) coordinates ($t,r,\theta,\phi)$,
\begin{align}
\ell &= \frac{r^2 + a^2}{\Delta}\partial_t + \partial_r + \frac{a}{\Delta}\partial_\phi, \\
n &= \frac{r^2 + a^2}{2\rho\rho^*}\partial_t - \frac{\Delta}{2\rho\rho^*}\partial_r + \frac{a}{2\rho\rho^*}\partial_\phi, \\
m &= \frac{1}{\sqrt{2}\rho}\left(ia \sin \theta \partial_t + \partial_\theta + \frac{i}{\sin \theta}\partial_\phi\right).
\end{align}
where a star denotes complex conjugation.  

The Kinnersley tetrad and BL coordinates are not regular on the event horizon.  For near-horizon analysis we will instead use ingoing coordinates $(v,r,\theta,\varphi)$ and the Hartle-Hawking tetrad $\{\hat{\ell},\hat{n},\hat{m},\hat{m}^*\}$.  The ingoing coordinates are defined by
\begin{align}
    v = t + r_*, \qquad \varphi=\phi+r_\sharp,
\end{align}
with $r_*$ and $r_\sharp$ satisfying
\begin{align}
    \frac{d r_*}{dr} = \frac{r^2+a^2}{\Delta}, \qquad \frac{dr_\sharp}{dr}=\frac{a}{\Delta}.
\end{align}
The Hartle-Hawking tetrad is defined by
\begin{align}
\hat{\ell} = B \ell, \ \ \
\hat{n} = B^{-1} n, \ \  \
\hat{m} = m
\end{align}
with boost
\begin{equation}
B = \frac{\Delta}{2(r^2+a^2)}.
\end{equation}

We use a hat to denote a quantity constructed in the regular tetrad and coordinates.  The hatted Weyl scalars satisfy
\begin{align}
\hat{\psi}_{s} = B^s \psi_s.
\end{align}
When considering Fourier mode solutions of the form
\begin{align}
    \psi_s = f(r,\theta) e^{i m \phi}e^{- i \omega t},
\end{align}
we will express the hatted version as 
\begin{align}
    \hat{\psi}_s = B^s \psi_s = \hat{f}(r,\theta) e^{i m \varphi}e^{-i \omega v},
\end{align}
so that 
\begin{align}
   \hat{f} = B^s e^{-i m r_\sharp}e^{i \omega r_*}f. 
\end{align}

We will commonly work with co-rotating perturbations with $\omega=m\Omega_H$.  In this case the relationship is
\begin{align}
    \hat{f} = B^s e^{-im(r_\sharp-\Omega_H r_*)}f.
\end{align}
This special combination is logarithmically divergent near the horizon,
\begin{align}\label{corotrelat}
    r_\sharp - \Omega_H r_* \sim - 2M\Omega_H\log x, \qquad x\to 0,
\end{align}
where $\sim$ denotes asymptotic equality and 
\begin{align}\label{xdef}
    x=\frac{r-M}{M}.
\end{align}
Thus we have
\begin{align}\label{fKN}
    \hat{f} \sim x^{2s+2iM\Omega_Hm} f,
\end{align}
which for the Kerr spacetime $(a=M)$ becomes
\begin{align}\label{fkerr}
    \hat{f} \sim x^{2s+im} f.
\end{align}

\section{Kerr black hole}\label{sec:Kerr}

We will start with the special case of a Kerr black hole ($Q=0$), where the mathematics simplifies significantly.  In this limit the gravitational and electromagnetic perturbations decouple,\footnote{In our convention for $\psi_{-1}$, a factor of $1/Q$ must be stripped off from \eqref{psim1} in order to obtain a finite Kerr limit.} controlled respectively by $\psi_{\pm2}$ and $\psi_{\pm1}$, which separately satisfy the spin-$s$ Teukolsky equation \cite{teukolsky1973perturbations}. 

A general perturbation may be Fourier transformed and expanded with respect to the spin-weighted spheroidal harmonics ${}_sS_{\ell m \omega}$.  Each individual ``term'' then has the form
\begin{align}
    {}_s\psi_{\ell m \omega} = {}_sR_{\ell m \omega}(r){}_sS_{\ell m \omega}(\theta) e^{i m \phi}e^{- i \omega t},
\end{align}
leading to an ODE for ${}_sR_{\ell m \omega}$,
\begin{align}\label{radialeqn}
    {}_s\mathcal{E}_{\ell m \omega}[{}_sR_{\ell m \omega}(r)]=0,
\end{align}
with the operator ${}_s\mathcal{E}_{\ell m \omega}$ given in Eq.~(28) of Ref.~\cite{Gralla:2017lto}. 
A co-rotating perturbation depends on $\phi$ and $t$ through the combination $\phi-\Omega_H t$ and hence has frequency $\omega=m\Omega_H$.  Such modes have been called ``synchronous'' in Ref.~\cite{Richartz:2017qep}.

It is easy to check that the Teukolsky equation at $\omega=m\Omega_H$ predicts near-horizon power-law behavior,
\begin{align}\label{psicorot}
    {}_sR_{\ell m (m\Omega_H)}\sim a x^{h-1-s} + bx^{-h-s}, \quad x\to0,
\end{align}
where $a$ and $b$ are constant coefficients, $x=(r-M)/M$ was defined in \eqref{xdef}, and the ``conformal weight'' $h$ is
\begin{align}\label{Kerrh}
    h = \frac{1}{2} + \sqrt{\frac{1}{4} + {}_sK_{lm} - 2m^2},
\end{align}
where ${}_sK_{lm}>0$ is the spheroidal harmonic eigenvalue evaluated at co-rotation  $\omega=m\Omega_H$.  (See e.g. Eq.~(24) of Ref.~\cite{Gralla:2017lto}.)  This eigenvalue is defined with the convention that $K = \ell(\ell+1)$ and hence $h = \ell+1$ in the axisymmetric case ($m=0$), which is also stationary on account of the co-rotating assumption.    The numerically-observed properties of  ${}_sK_{lm}$  indicate that $h$ subdivides into three cases (see e.g. (67) of Ref.~\cite{Gralla:2017lto}),
\begin{align} \label{h_cases}
h \textrm{ is} \begin{cases} 
1/2 + i\delta, & |m| \gtrsim 0.74 l \quad \quad \ \!\text{(principal)}, \\ 
> 1, & 0 < |m| \lesssim 0.74 l \  \text{(supplementary)}, \\ 
l + 1, & m = 0 \quad \quad \quad \quad\ \text{(axisymmetric)},
\end{cases}
\end{align}
where $\delta >0$.  The transition from supplementary to principal at $m=0.74\ell$ is exact in the large-$\ell$ limit \cite{Yang:2013uba}.

In the horizon-regular Hartle-Hawking tetrad, the two power laws are instead
\begin{align}\label{psicorotHH}
    {}_s\hat{R}_{\ell m (m\Omega_H)}\sim ax^{h-1+s+im} + b x^{-h+s+im}
\end{align}
according to \eqref{fkerr}.   In the axisymmetric case $m=0$, the conformal weight $h=\ell+1$ is always a positive integer larger than $|s|$, meaning that the $a$ term is always regular, while the $b$ term is always singular.  When considering stationary, axisymmetric sources, one therefore imposes $b=0$ as a boundary condition.

The same approach does \textit{not} work for a truly co-rotating perturbation $m\neq 0$.  For the supplementary modes $h>1$, the $a$ term is still more regular than the $b$ term, but both terms fail to be smooth because of the non-integer power law.  The situation is even worse for the principal modes $h=1/2+i \delta$, where both terms behave as $x^{-1/2+s}$ up to a phase.  The terms are indistinguishable based on regularity, and both diverge for $s\leq 0$.

Horizon regularity is therefore not a reasonable boundary condition for the co-rotating modes.  To understand the situation we will need to make two changes.  First, we will consider modes that are \textit{nearly} co-rotating,
\begin{align}
    k \equiv 2\frac{\omega-m\Omega_H}{\Omega_H} \ll 1,
\end{align}
defined with a conventional factor of 2 \cite{Gralla:2017lto}.  Second, we will explicitly include a source,
\begin{align}\label{radialeqnSource}
    {}_s\mathcal{E}_{\ell m \omega}[{}_sg_{\ell m \omega}(r,r')]= \Delta^{-s} \delta(r-r'),
\end{align}
again with a conventional prefactor $\Delta^{-s}$ as in Eq.~(32) of Ref.~\cite{Gralla:2017lto}.  With these two changes, we may impose the standard boundary conditions of ingoing-wave behavior at the horizon, and outgoing-wave behavior at infinity.  In this context, ${}_sg_{\ell m \omega}(r)$ are interpreted as the modes of the retarded Green function for the Teukolsky equation.  

The solution for ${}_s g_{\ell m \omega}$ may be constructed explicitly using the method of matched asymptotic expansions (e.g., \cite{Gralla:2017lto}), considering separately the ``near'' approximation $x \ll 1$ and ``far'' approximation $x \gg k$ before matching in the region of overlap $k \ll x \ll 1$.  The near approximation is used to impose ingoing-wave behavior at the event horizon, and the far approximation is used to impose outgoing-wave behavior at infinity.  The power laws \eqref{psicorot} appear in the overlap region, where near and far are matched together. 

In order to understand the excitation of the overlap-region power laws by external sources, we choose the source point to be in the far region ($x'\gg k$) and the field point to be in the overlap region ($k\ll x\ll1$).  The non-axisymmetric Green function modes are then given by expanding Eq.~(54) of Ref.~\cite{Gralla:2017lto} in the overlap region, 
\begin{align} \label{transfer_overlap}
{}_s&g_{lm\omega}(x,x')  \approx  \frac{R(x')}{1-2h}\frac{x^{h-1-s} + \eta (-ik)^{2h-1} x^{-h-s} }{1 - \mathcal{R} \eta (-ik)^{2h-1}},
\end{align}
where $R(x')$ is a far-zone solution (called $R^{\rm far, up}$ in Ref.~\cite{Gralla:2017lto}), and
\begin{align}
\eta
&= -\,\frac{\Gamma(2-2h)\,\Gamma(h-im-s)}
{\Gamma(2h)\,\Gamma(1-h-im-s)}, \\[1ex]
\mathcal R
&= -\,\frac{\Gamma(2-2h)\,\Gamma(h-im+s)}
{\Gamma(1-h-im+s)\,\Gamma(2h)}
(-im)^{2h-1}.
\end{align}

In the limit $k \to 0$, we may compare with the precisely co-rotating behavior \eqref{psicorot}.  For a supplementary (or axisymmetric) mode, we have $2h-1>0$ so that $k^{2h-1}\to 0$ as $k\to 0$, and we recover just the $a$ term of \eqref{psicorot}, showing that $b=0$ is indeed the correct boundary condition for these modes.  However, for a principal mode, $2h-1=2i\delta$ is purely imaginary, and $k^{2h-1}$ is an infinitely oscillating phase as $k\to0$.  
  
For these modes we cannot set $k=0$ precisely and instead must regard the factors of $k^{2h-1}$ as a ``random phase'' for small but non-zero values of $k$.  In other words, both terms of \eqref{psicorot} contribute as $k \to 0$ for nearly co-rotating sources, with coefficients that oscillate with $k$ (but remain bounded) according to \eqref{transfer_overlap}.  Therefore, the perturbation is of order $x^{-1/2-s}$ when $k \ll x \ll 1$.  

Using Eq.~\eqref{fkerr}, we see that the regular-tetrad Green function modes $\hat{g}$ are of order $x^{-1/2+s}$ in the overlap region $k \ll x \ll 1$.  This indicates arbitrarily large $s\leq 0$ perturbations as $k\to0$.  Evaluating at $x \sim k^{p}$ for $0<p<1$, the perturbation has size
\begin{align}
    \hat{g}(x \sim k^p) \sim k^{p(-1/2+s)}, \qquad k \to 0.
\end{align}

To understand the physical effects, we use the estimate $x^{-1/2+s}$ for the magnitude of the perturbation with a cutoff at $x\sim k$ where the overlap region ends.  For $s=-1$ we have a component of electric field of order $x^{-3/2}$ experienced by an infalling observer.   Since $m^\mu$ represents angular directions, from \eqref{psim1} we see that the leading singular electric (and magnetic) field is  transverse to the horizon.  The energy $\Delta E$ imparted to the observer scales as one integral of the behavior, i.e. $x^{-1/2} \sim k^{-1/2}$, and hence is large:
\begin{align}
        \Delta E &\sim k^{-1/2} \to \infty \qquad (\textrm{electromagnetic}).
\end{align}
However, the contribution to the integrated displacement goes as $x^{1/2}\sim k^{1/2}$ and is small.  In other words, the body experiences a large kick transverse to the horizon, while the angular displacement remains finite.

For $s=-2$ the perturbation represents a component of perturbed spacetime curvature of order $x^{-5/2}$ experienced by an infalling observer.  Again from \eqref{psim2} we see that the leading tidal field is transverse.  The tidal deformation $\Delta \xi$ scales as two integrals of this behavior, giving $x^{-1/2}\sim k^{-1/2}$,
\begin{align}
    \Delta \xi &\sim k^{-1/2} \to \infty \qquad (\textrm{gravitational}).
\end{align}
In other words, the infalling body experiences large tidal deformation transverse to the horizon.  It also experiences a large rotation transverse to the horizon (though only logarithmically so), stemming from the imaginary part of the principal exponent $x^{-1/2+i \alpha}$.  

We emphasize that claims of \textit{arbitrarily} large deformation apply only within linearized theory.  In fact, it seems clear that the metric perturbation itself should scale in the same way as the tidal deformation, such that arbitrarily large  deformation just means that linearized theory is breaking down.  Within linearized theory, the proper interpretation of ``large'' tidal deformation is \textit{larger than expected}.  As an observer falls toward a perturbed Kerr black hole under normal circumstances, he experiences a gradual increase in tidal force set by the typical scales of the background spacetime, with no special sign of horizon-crossing.  However, if the perturbation is nearly co-rotating, the observer instead feels a sudden increase in tidal force right near the horizon.  More precisely, if $\epsilon$ represents the typical strength of the perturbation far from the black hole, the observer experiences deformation of order  $\epsilon k^{-1/2}$ over the last moments of his journey passing through the region $k \ll x\ll1$.

We now discuss how co-rotating modes can be excited by a persistent source.  (Although we will say ``co-rotating'' for short, we always have in mind a limiting process involving a nearly co-rotating mode.) The retarded solution for $\psi_s$ resulting from a physical source (conserved stress-energy or charge-current) is constructed by applying certain differential operators \cite{teukolsky1973perturbations} to the stress-energy or charge-current and then integrating against the Teukolsky equation Green function that we study.  These operators do not mix Fourier modes, so we may focus on the co-rotating modes of the source.  We restrict consideration to sources having a discrete set of non-zero Fourier modes that include co-rotating frequencies.  The simplest example is a co-rotating particle, i.e. a particle following an orbit of $\pd_t+\Omega_H \pd_\phi$ in a region where that vector is timelike.  For electromagnetic perturbations, the particle need not follow a geodesic, and we may simply place it sufficiently close to the symmetry axis at any given ``height''.  For gravitational perturbations, however, the source must follow a geodesic, and the extremal Kerr spacetime has no co-rotating timelike geodesics.  In particular, $\pd_t+\Omega_H \pd_\phi$ is everywhere spacelike on the equatorial plane outside the horizon.

However, we may instead consider eccentric and/or inclined orbits that are tuned to have co-rotating harmonics.  The key point is that bound orbits are characterized by three frequencies $\Omega_r,\Omega_\theta,\Omega_\phi$ associated respectively with radial, polar, and azimuthal oscillations \cite{Carter:1968rr,Schmidt:2002qk}.  As a consequence, any scalar covariantly constructed from the orbit along with other stationary, axisymmetric tensors will have time dependence given by a sum over frequencies
\begin{align}\label{omegamnp}
    \omega_{mnq} = m \Omega_\phi + n \Omega_r + q \Omega_\theta,
\end{align}
where $m,n,q$ are integers.  A harmonic is co-rotating when $\omega=m \Omega_H$, or
\begin{align}\label{corotres}
    n \Omega_r + q \Omega_\theta + m (\Omega_\phi-\Omega_H)=0.
\end{align}

These resonances can be obtained by tuning a single orbital parameter. For example, we parameterize an equatorial orbit by
a dimensionless radial parameter $p_{\mathrm{orb}}$ and eccentricity $e$, defined by the radial turning points
$r_{\min}/M=p_{\mathrm{orb}}/(1+e)$ and
$r_{\max}/M=p_{\mathrm{orb}}/(1-e)$. Setting $e=0.1$, we find that the resonance condition
\begin{equation}\label{kerrRes}
    \Omega_\phi+3\Omega_r=\Omega_H
\end{equation}
is satisfied at $p_{\mathrm{orb}}=1.68854$, where $M\Omega_r=0.0619771$ and $M\Omega_\phi=0.314069$. This orbit has specific energy $\mathcal{E}=E/\mu=0.741171$ and specific angular momentum $\mathcal{L}=L_z/\mu=1.55742M$, and oscillates between $r_{\min}=1.53504M$ and $r_{\max}=1.87616M$.

The $q=0, n=3m$ harmonics of this orbit are co-rotating \eqref{corotres}.  In particular, the $m=2$ harmonic
\begin{equation}
    \omega_{260}=2\Omega_\phi+6\Omega_r=2\Omega_H
\end{equation}
contains the $\ell=m=2$ dominant principal mode and associated singular horizon behavior.  In Appendix~\ref{app:resonant-vlasov-source} we show how one can manipulate orbital phases of a large number of particles in order to cancel off the non co-rotating harmonics and create a precisely co-rotating source.

\section{Kerr-Newman Black hole}\label{sec:KN}

We now consider the general case of non-zero black hole charge, $Q\neq 0$.  The coupled electromagnetic and gravitational perturbations are encoded in $\psi_{-1}$ and $\psi_{-2}$ \cite{chandrasekhar1998mathematical,dias2015linear,mark2015quasinormal}.  Each mode $\psi_i(r,\theta) e^{i m \phi}e^{- i \omega t}$ 
satisfies the coupled equations 
\begin{align}
\left(F_{-2} + Q^2 G_{-2} \right)\psi_{-2} + Q^2 H_{-2}\psi_{-1} &= 0,
\label{eq:PDE1}\\
\left(F_{-1} + Q^2 G_{-1} \right)\psi_{-1} + Q^2 H_{-1}\psi_{-2} &= 0,
\label{eq:PDE2}
\end{align}
where $F_i,G_i,H_i$ are second-order differential operators in $r$ and $\theta$.  (We use the notation of Ref.~\cite{dias2015linear}, where explicit expressions may be found.)  In this section, $\psi_{s}(r,\theta)$ refers to the radial and polar dependence of a mode; we suppress dependence on $m$ and $\omega$.

Although some aspects of these equations are amenable to analytic analysis~\cite{DiasGodazgarSantos2022}, in general they must be solved numerically.  We will proceed in two steps, mirroring our analytical analysis of the Kerr case.  First we will assume precise co-rotation ($k=0$) and seek special solutions with power-law dependence in $x$ as $x\to0$.  At each value of black hole charge $Q$ and azimuthal number $m$, we find a discrete, apparently infinite, set of solutions that generalize the mode-by-mode power laws of Kerr \eqref{psicorot}, to which they reduce in the zero-charge limit.  We then confirm these exponents arise in physical solutions by numerically constructing the retarded Green function at small-but-finite values of $k$.

\subsection{Critical exponents of the KN spacetime}\label{sec:KN-exponents}

To identify the critical exponents of the KN spacetime we set $k=0$ ($\omega=m\Omega_H$) and make the ansatz
\begin{align}\label{NHansatz}
    \psi_{s}(x,\theta)=x^{w-s} S_s(\theta) \quad \textrm{as } x\to 0,
\end{align}
where $x=(r-M)/M$ as before.  (The number $w$ is the invariant weight of the perturbation, defined in Sec.~\ref{sec:weight} below.) This yields a pair of purely angular second-order ordinary differential equations of the form
\begin{align}\label{angularode1}
\mathcal{L}_{-2}(\theta;m,w)[S_{-2},S_{-1}] &= 0,\\
\mathcal{L}_{-1}(\theta;m,w)[S_{-2},S_{-1}] &= 0,\label{angularode2}
\end{align} 
which depend on the weight $w$.  We expect based on the Kerr analysis that pole-regular solutions will be possible only for certain choices of weight.  In the precise Kerr limit $Q=0$, the allowed weights are $w=h-1$ and $w=-h$ with $h$ given by \eqref{Kerrh}, and the equations decouple into separate spheroidal harmonic equations for each value of $s$.  Away from $Q=0$, the equations are coupled, but we still expect a discrete set of pole-regular solutions labeled by discrete allowed choices of $w$.

\begin{figure}
    \centering
    \includegraphics[width=1\linewidth]{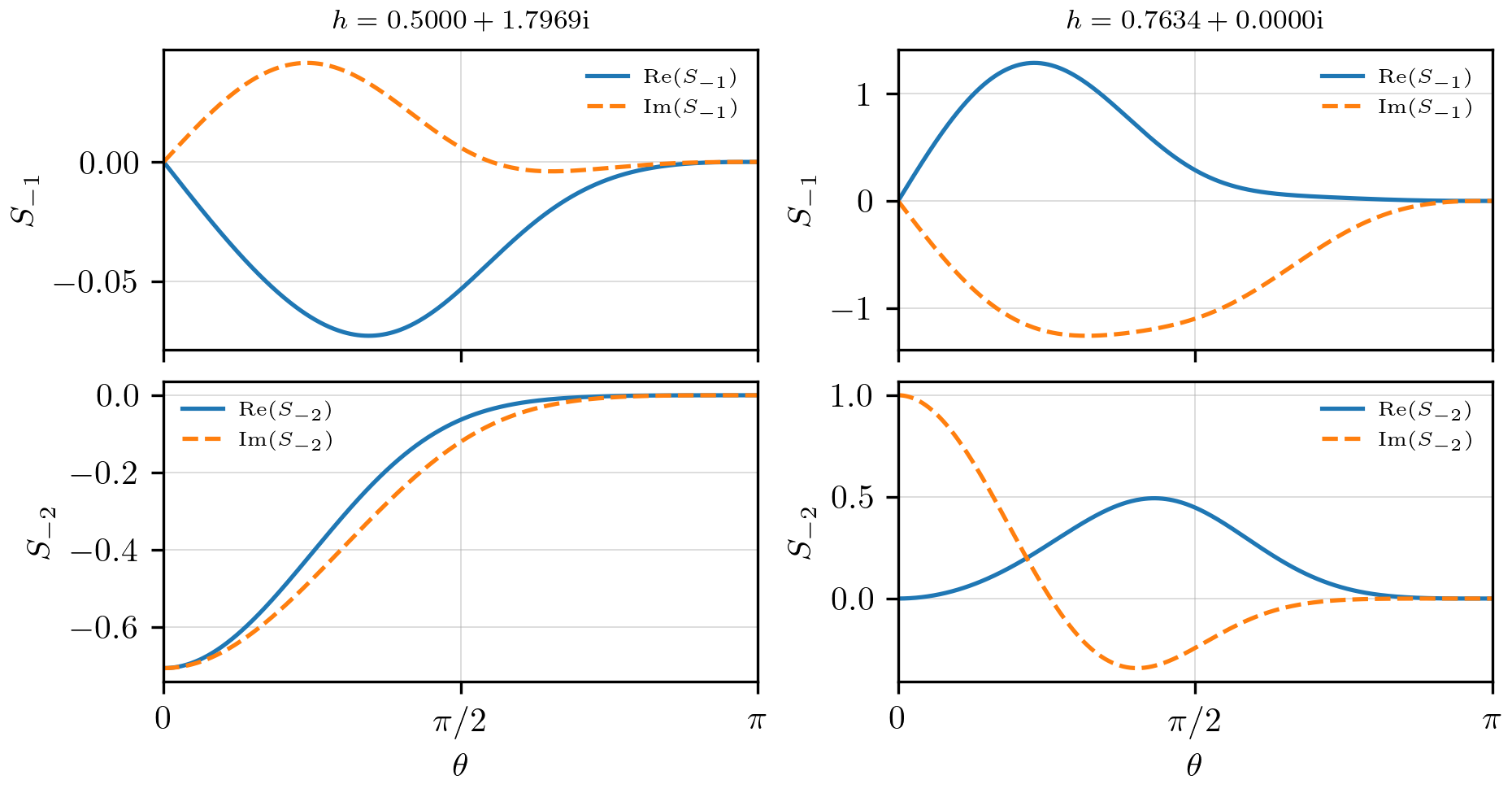}
    \caption{Angular dependence of near-horizon self-similar perturbations of the extremal KN black hole.  We take $Q/M=.67$ and $m=2$ and show the lowest two modes, with overall normalization chosen for convenience. These functions reduce to spin-weighted spheroidal harmonics in the Kerr limit. }
\label{fig:KN_eigenfunctions}
\end{figure}

To find these solutions we adopt a spectral method, where pole-regularity is imposed at the grid level.  This rephrases Eqs.~\eqref{angularode1}--\eqref{angularode2} as a generalized matrix eigenvalue problem, with $w$ the eigenvalue. The detailed method is described in Appendix \ref{app:numerical_exp}, and some results are shown in Fig.~\ref{fig:KN_eigenfunctions}.  We validate our method with convergence tests and by reproducing the spin-weighted spheroidal harmonics as $Q \to 0$, where the KN metric approaches the Kerr metric.

\begin{figure}
    \centering
    \includegraphics[width=1\linewidth]{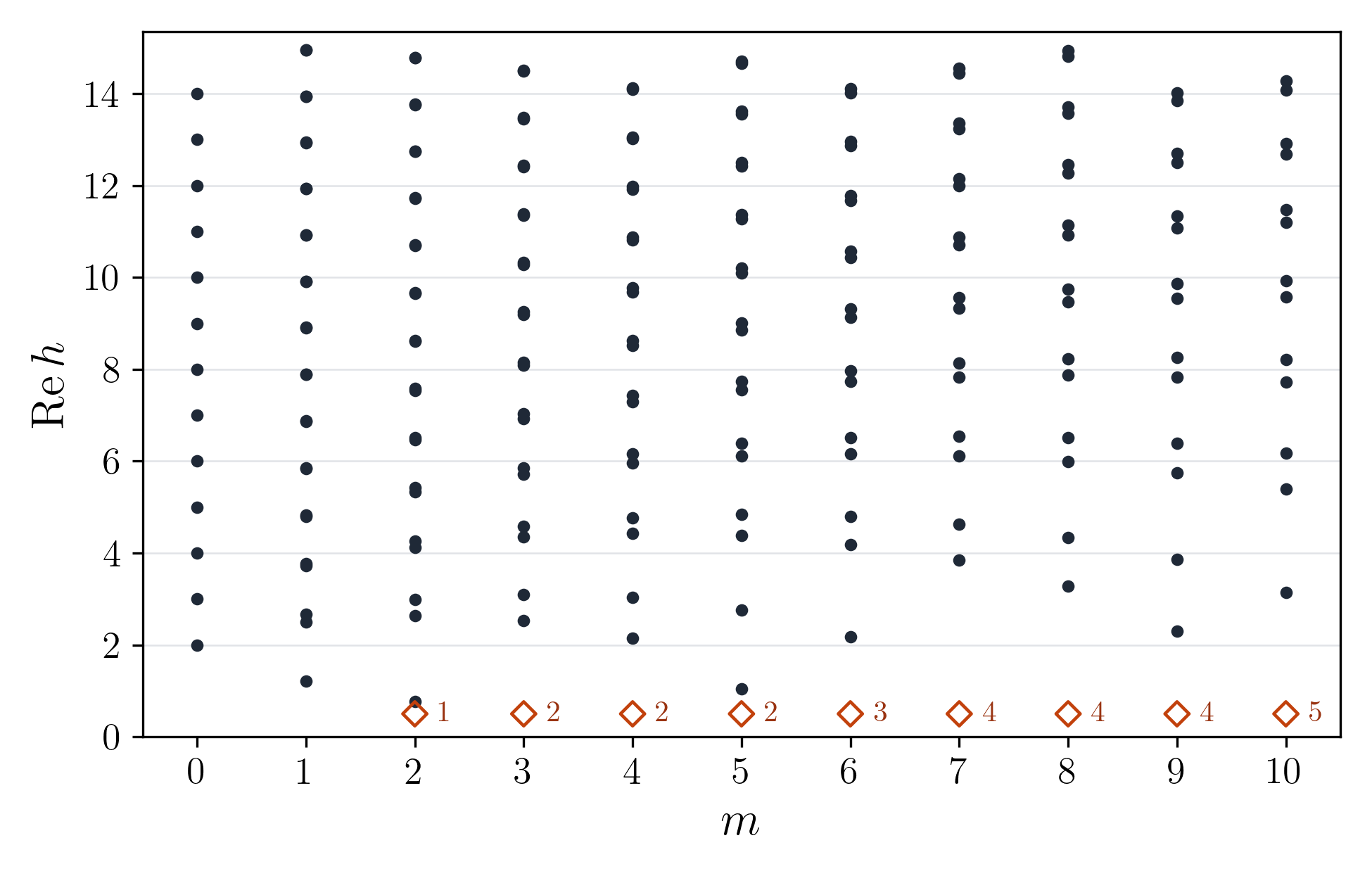}
    \caption{Conformal weights $h$ of KN perturbations at $Q/M=0.67$. Black circles denote supplementary modes (real $h$), while red diamonds denote principal modes $h=1/2+i\delta$, with adjacent numbers giving their multiplicity (distinct values of $\delta>0$). 
    The  numerical values are given in 
    Tab.~\ref{tab:kn-physical-h-grid} below.  At small enough $m$, the conformal weights appear in visible pairs that coalesce at $m=0$. In the Kerr limit where the spin sectors decouple, each member of a pair corresponds to one value of $s=\{-1,-2\}$ according to Eq.~\eqref{Kerrh}.}
    \label{fig:Q=.67}
\end{figure}

We find that the exponents $w$ take a very similar form to the Kerr metric.  They are either real numbers or complex numbers with real part equal to $-1/2$.   Furthermore, they always appear in pairs $w \leftrightarrow -1-w$.  To represent these pairs analogously to the Kerr spacetime \eqref{psicorot}, we write each pair of weights in terms of a complex number $h$ as
\begin{align}\label{wh}
    w=\{h-1,-h\}
\end{align}
with the convention that $h>0$ when real, and $\textrm{Im}[h]>0$ when complex (in which case $\textrm{Re}[h]=1/2$), and $m\geq0$.\footnote{The 
coupled equations possess the symmetry
$
 (\psi_{-2},\psi_{-1})_{-m,-\omega^*}(r,\theta)
 =
 \bigl(\psi_{-2}^*,-\psi_{-1}^*\bigr)_{m,\omega}(r,\pi-\theta).
$
For synchronous modes $\omega=m\Omega_H$, this sends $m \to -m$ and $w\to w^*$; supplementary weights
are therefore unchanged, while principal weights are conjugated.}  We can therefore represent all the weights by listing the values of $h$ for $m\geq 0$, which all take the form $h=1/2+\sqrt{C}$ for real $C$, analogously to \eqref{Kerrh} for Kerr.  Examples are shown in Fig.~\ref{fig:Q=.67} and Tab.~\ref{tab:kn-physical-h-grid}.  In order to distinguish these two notions of weight, we refer to $w$ as the \textit{tensor weight} and $h$ as the \textit{conformal weight}.

In the axisymmetric case $m=0$, we find that the conformal weights are positive integers starting with $2$,
\begin{align}\label{axih}
    h = \mathbb{Z}_{\geq 2} \qquad \textrm{($m=0$)}.
\end{align}
This is consistent with the Kerr case \eqref{Kerrh} where $h=\ell+1$ is known analytically, noting that the lowest value for electromagnetic perturbations is $\ell=1$.  It is also consistent with the analytic solution presented in Sec.~3 of Ref.~\cite{Horowitz:2024dch}, as explained in Sec.~\ref{sec:weight-comparison} below.  The fact that GEM perturbations of the KN spacetime start with effective $\ell=1$ can be attributed to the coupling between the electromagnetic and gravitational sectors.

\begin{figure}[t]
    \centering
    \includegraphics[width=1\linewidth]{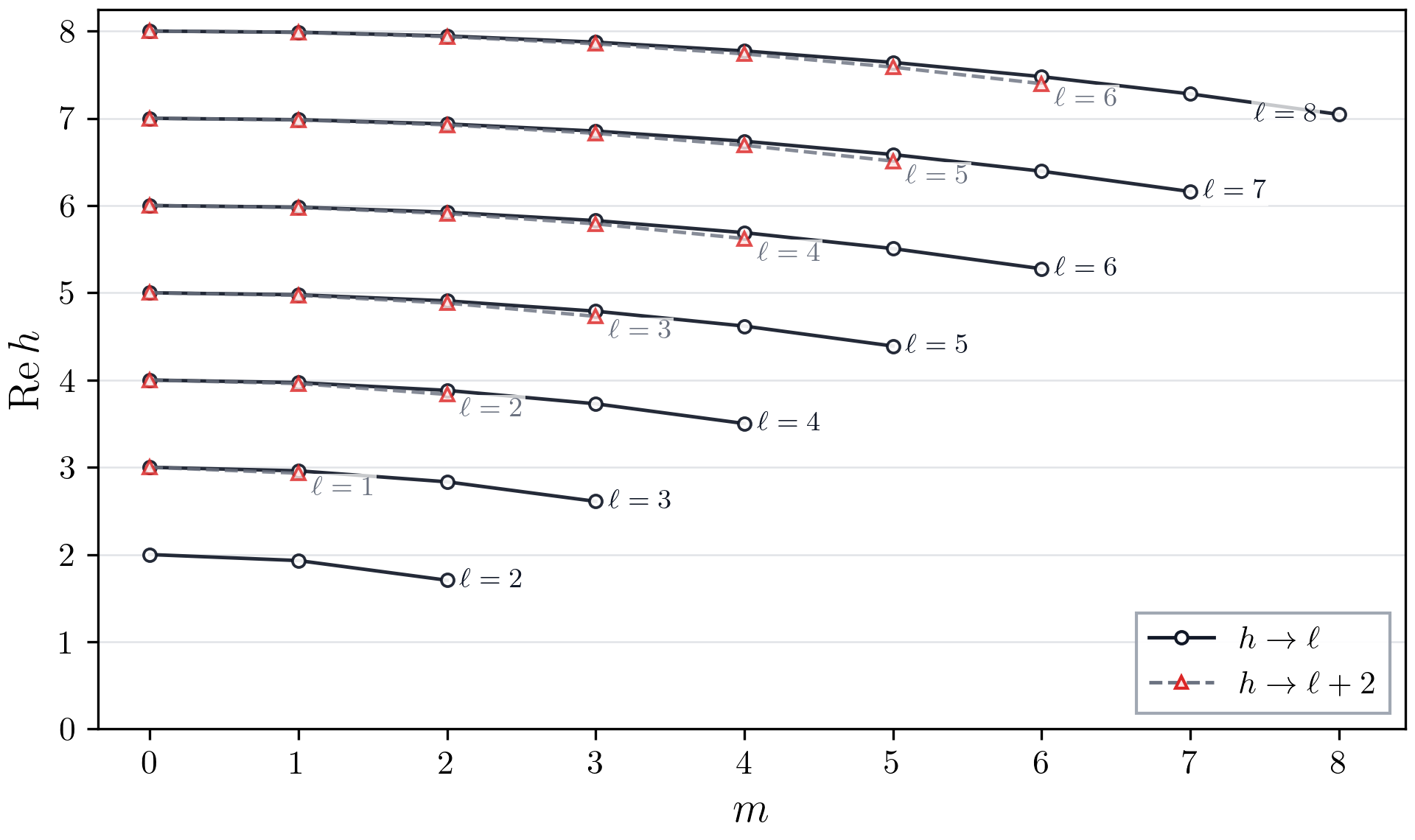}
    \caption{Degeneracy of conformal weights $h$ near the RN limit $Q \to M$.  Here we show the weights for $Q/M=0.98$, where the two branches can be visually distinguished.  Each weight is assigned to the $\ell$ branch or the $\ell+2$ branch by its continuation to the RN limit $Q/M=1$, where 
the angular eigenfunctions reduce to spin-weighted spherical harmonics and carry a well-defined value of $\ell$.}
    \label{fig:RN}
\end{figure}

In the RN limit $Q\to M$, we find that the weights become integers beginning with the larger of $m$ and $2$,
\begin{align}\label{RNh}
    h= \mathbb{Z}_{\geq \textrm{max}(2,m)} \qquad \textrm{(RN spacetime)}.
\end{align}
This curious structure can be explained by labeling the weights according to their corresponding representation of $SO(3)$.  That is, since the background is spherically symmetric, the eigenfunctions $\{S_{-1},S_{-2}\}$ reduce to spin-weighted spherical harmonics that carry a definite value of $\ell$.  We find that there are two distinct families of conformal weights for $Q$ near $M$, labeled by their distinct behaviors $h\to \ell$ and $h\to \ell+2$ as $Q \to M$.  The first family begins at $\ell=2$, and the second family begins at $\ell=1$. In particular, the modes with $h \geq 3$ and $h -m \geq 2$ are doubly degenerate, while the remaining modes are non-degenerate.  These families are illustrated in Fig.~\ref{fig:RN}.  This structure is also in precise agreement with the analytic RN solution of \cite{Porfyriadis:2018yag}, as explained in Sec.~\ref{sec:weight-comparison} below.

\begin{table*}[p]
\centering
\scriptsize
\setlength{\tabcolsep}{5pt}
\renewcommand{\arraystretch}{1.02}
\caption{Representative conformal weights of co-rotating GEM perturbations of extremal KN spacetime.  A dash means that no mode of that type was found.  Each azimuthal mode has a finite (possibly empty) set of principal modes together with an infinite number of supplementary modes.  The axisymmetric $m=0$ weights (not shown) are $2,3,4,5,\dots$ at all values of charge.}
\label{tab:kn-physical-h-grid}
\includegraphics[page=1,width=\textwidth,trim=70 67 70 125,clip]{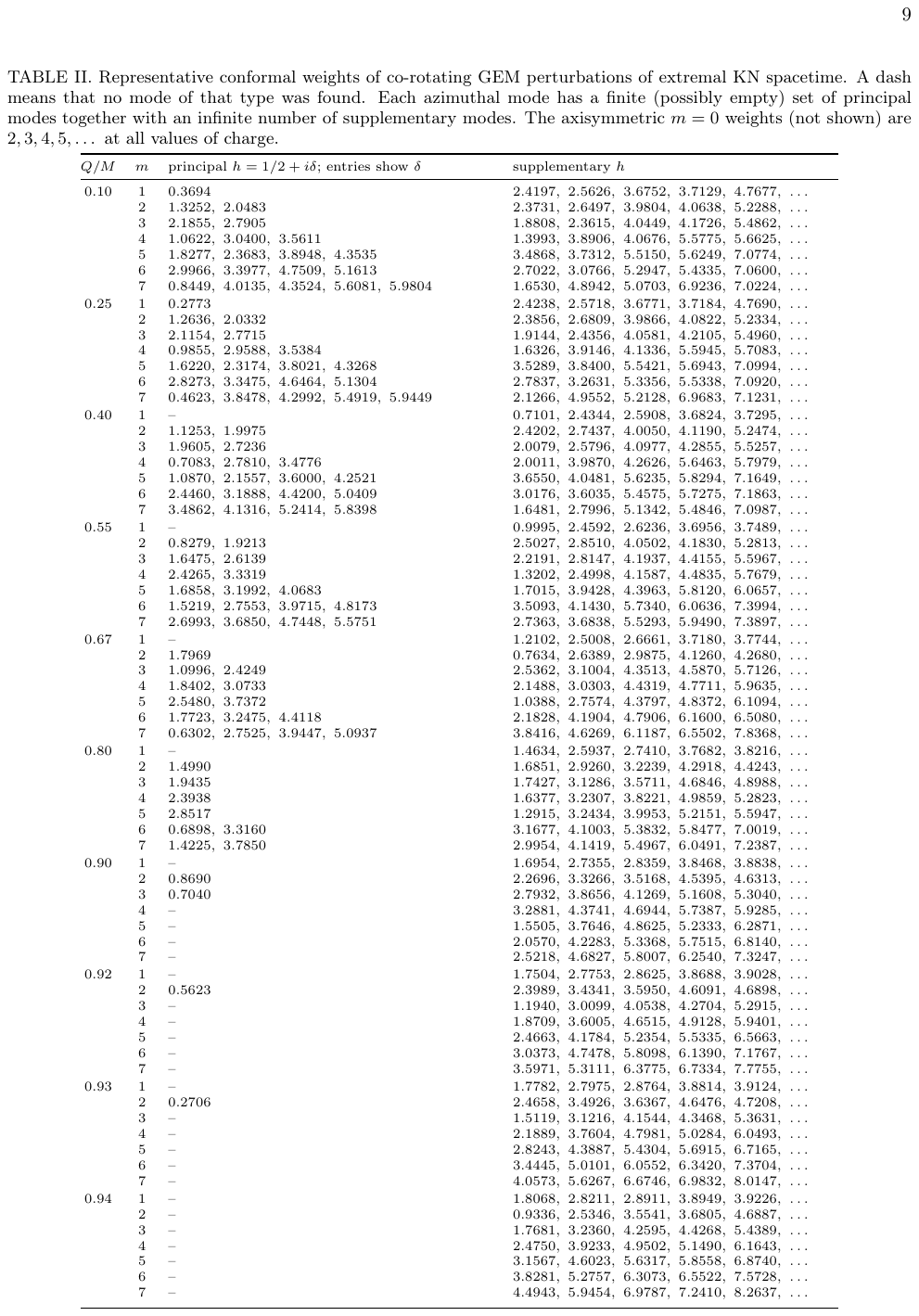}
\end{table*}

\begin{figure}
    \centering
    \includegraphics[width=1\linewidth]{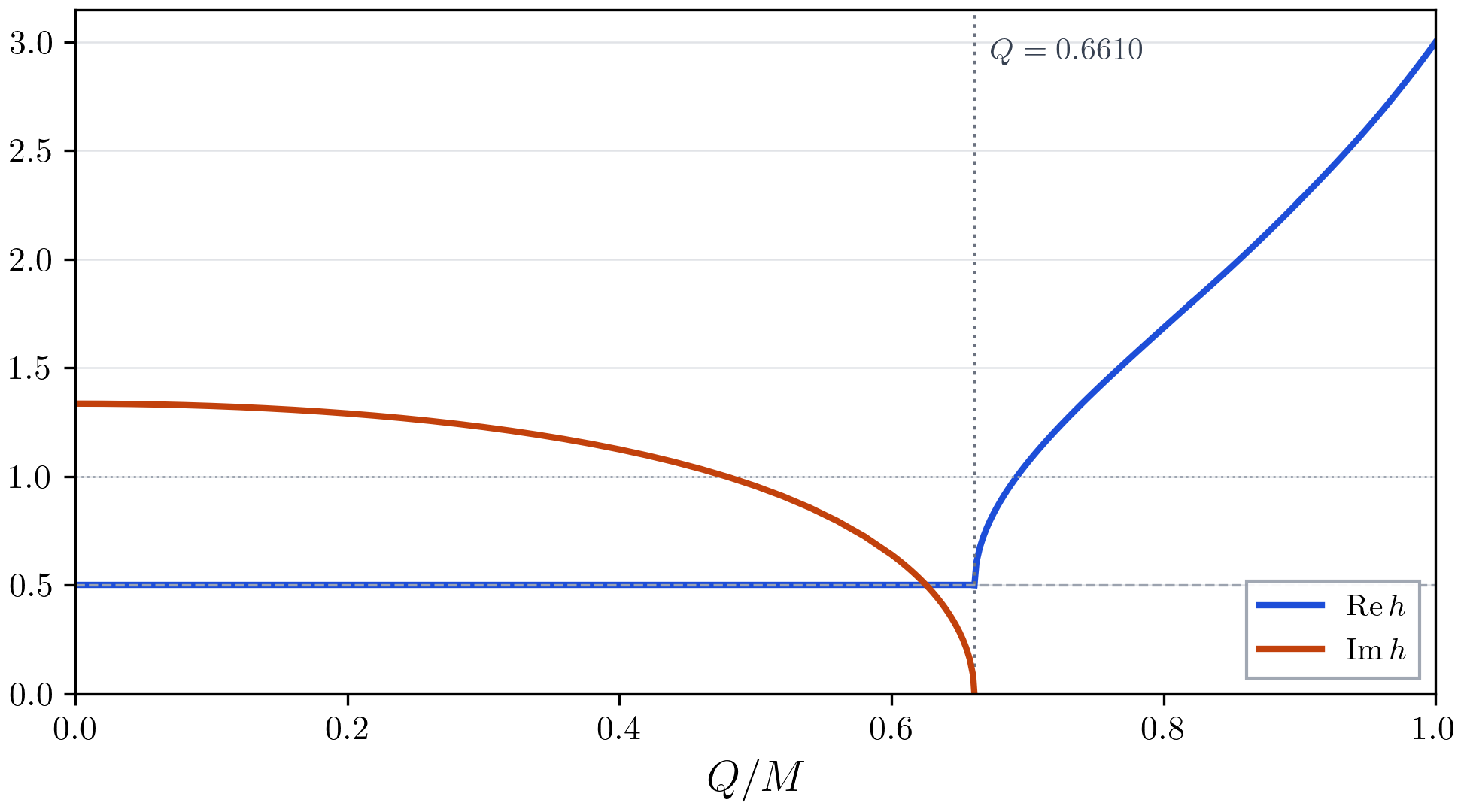}
    \caption{Conformal weight $h$ of an $m=2$ mode as a function of black hole charge $Q$. The mode begins as a principal mode at $h=\tfrac12+1.33609i$ at $Q=0$ (Kerr spacetime), changes to supplementary at  $Q\approx0.66096$, and approaches the $h=3$ at $Q=M$ (RN spacetime).}
    \label{fig:bifurcation}
\end{figure}

\begin{figure}[t]
\centering
\includegraphics[width=1.0\linewidth]{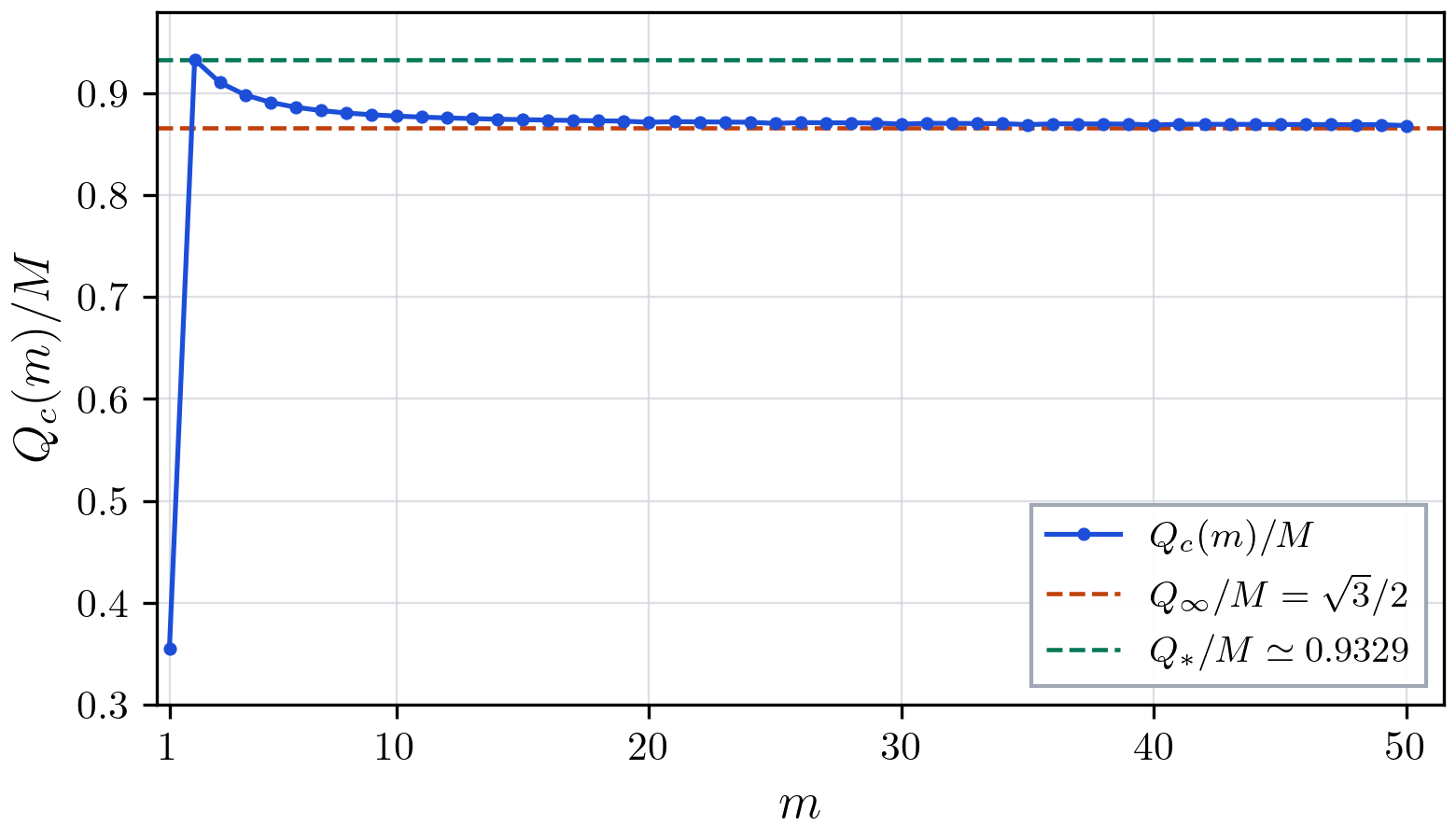}
\caption{Upper bound on $Q$ for the existence of principal modes, as a function of azimuthal number $m$.  For $Q<Q_c(m)$, the spectrum contains principal modes $h=\frac12+ i\delta$.  The maximum occurs at $m=2$, where $Q=Q_{*}\approx0.9329M$, while at large $m$ the curve approaches $Q_\infty=\sqrt{3}/2\approx 0.8660M$.
}
\label{fig:Qc_vs_m}
\end{figure}

\begin{figure}
    \centering
    \includegraphics[width=1\linewidth]{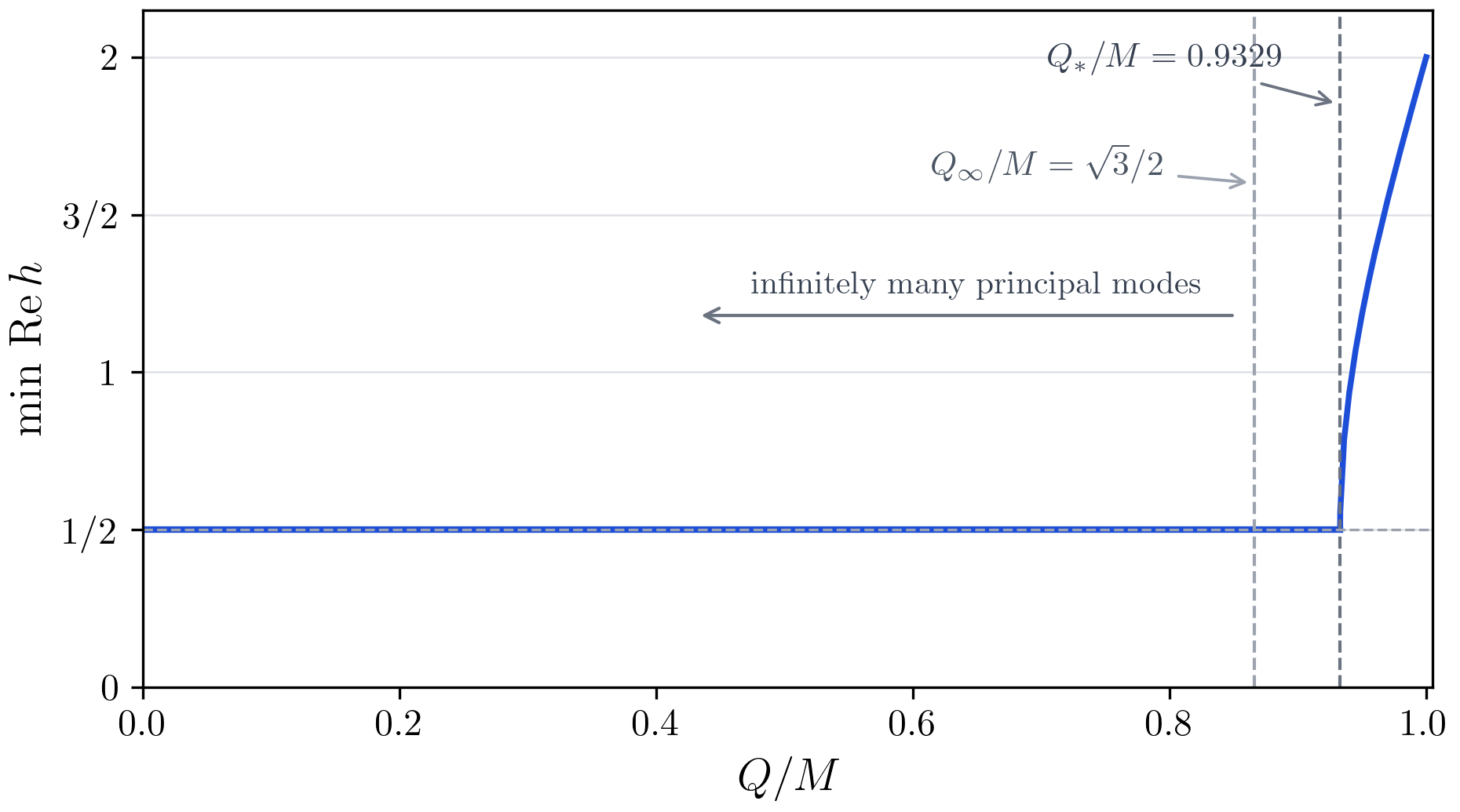}
    \caption{Smallest real part of conformal weight as a function of black hole charge. For $Q<Q_\infty=(\sqrt{3}/2)M$ the spectrum contains infinitely many principal modes $h=1/2+i\delta$, while for $Q_\infty<Q<Q_{*}\approx0.9329M$ the spectrum contains a finite number of principal modes. Above this critical value the
principal sector is absent and the minimum $h$ is supplied by the real $m=2$
branch.  In the RN limit $Q\to M$ the smallest weight is $h=2$ exactly.}
    \label{fig:minh}
\end{figure}

We can also follow individual exponents as a function of the charge $Q$ of the black hole.  The Kerr spacetime ($Q=0$) has principal modes, while the RN spacetime $Q=M$ does not.  Thus if we begin with a principal mode at $Q=0$ and increase $Q$, then it will eventually become supplementary at a particular value (Fig.~\ref{fig:bifurcation}). For each $m$, the largest such value among all modes will be called the critical charge and denoted $Q_c(m)$.  Plotting this critical charge as a function of azimuthal number $m$  (Fig.~\ref{fig:Qc_vs_m}) reveals two important thresholds.  First, the largest critical charge occurs for $m=2$ and has value
\begin{align}\label{Qstar}
    Q_* \approx 0.9329M,
\end{align}
in agreement with prior calculations (see text below equation (3.41) of Ref.~\cite{DiasGodazgarSantos2022}).
Second, the critical charge approaches a constant value as $m\to\infty$,
\begin{align}\label{Qinfty}
    Q_\infty = \frac{\sqrt{3}}{2}M \approx .8660M.
\end{align}
(The exact value $\sqrt{3}/2$ follows from the WKB analysis of Ref.~\cite{DiasGodazgarSantos2022}, as we describe in Sec.~\ref{sec:weight-comparison} below.)  These values encode the number and existence of principal modes:
\begin{align}
    & Q < Q_\infty: \quad \textrm{Infinitely many} \label{infinite} \\
    Q_\infty < \ & Q < Q_*: \quad  \ \textrm{Finite number} \label{finite} \\
    & Q > Q_*: \quad  \ \textrm{None} \label{none} 
\end{align}
Fig.~\ref{fig:minh} illustrates this transition.

The tidal deformation caused by a mode scales as $k^{h-1}$, so supplementary modes with $h<1$ still have infinite tidal deformation.  We find that the range for the existence of a supplementary mode with $h<1$ is $Q \lesssim 0.94227M$.  Modes with $h=1/2$ exactly are expected to display distinctive behavior \cite{Gajic:2026mre}.  We do not study this special case here.

\subsection{KN Green function near co-rotation}\label{sec:KNgreens}

In the previous subsection we computed the allowed KN weights $w$ under a near-horizon scaling ansatz \eqref{NHansatz}.  We found that the weights appear in pairs which are indexed by a single ``conformal weight'' $h$ as $w=h-1$ and $w=-h$, just as in the Kerr spacetime.  In the Kerr spacetime, the analytical analysis of the retarded Green function near co-rotation  \eqref{transfer_overlap} established that only the $w=h-1$ branch is relevant for supplementary modes, while both branches are relevant for principal modes (having the same real part $\textrm{Re}[w]=-1/2$). In App.~\ref{app:green} we describe a numerical method that confirms the same behavior in the KN case.  The basic idea is to use a generic spectral method to find the retarded Green function at a numerically small value of $k$ and then fit power-law behavior in the overlap region $k \ll x\ll 1$. This confirms the presence of singular horizons for GEM perturbations of the KN spacetime.

\subsection{Excitation of co-rotating perturbations}

Our discussion of the excitation of co-rotating perturbations in the Kerr spacetime relied on eccentric and/or inclined orbits.  We expect that a similar construction can be performed for the KN spacetime at any value of charge, but here we note the simpler option of a co-rotating equatorial particle above a threshold value of charge.

In the KN spacetime, the range of radii on the equatorial plane at which the Killing vector $\chi=\partial_{t}+\Omega_H\partial_{\phi}$ is timelike is 
\begin{equation}
    1 < \frac{r}{M} < \frac{1}{\sqrt{1 - (Q/M)^2}} - 1 .
\end{equation}
This range is non-empty as long as $Q/M>\sqrt{3}/2$.

A conserved point-particle source must obey the Lorentz force law.  One can choose the charge-to-mass ratio of the particle to ensure that an orbit is possible at any given value of $r$ in this range.  It is easy to check that the needed ratio is
\begin{equation}
    \frac{e}{\mu}
    =
    \frac{Q^{2} M^{3} - a^{2} r \left( r^{2} + M r + M^{2} \right)}
         {Q M^{2} \sqrt{M^{4} - a^{2} (r + M)^{2}}}.
\end{equation}
The needed charge-to-mass diverges at the upper end of the range.  In the RN limit  $Q\to M$ we have the expected gravito-electric balance $|e/\mu|\to1$.

In Sec.~\ref{sec:KN-exponents} we found that principal modes (and hence singular horizons) occur in the range $Q<Q_* \approx 0.93M$ \eqref{Qstar}.  Co-rotating timelike sources exist for $Q/M>\sqrt{3}/2\approx0.87$.  Over this range of $Q$, the co-rotating orbits are rather close to the horizon.  For example, at $Q/M=0.90$, co-rotating orbits are confined to $r\lesssim 1.3M$.  At $r=1.2M$, the charge-to-mass ratio is $|e/\mu|\approx0.078$.

\section{Tensor weight interpretation}\label{sec:weight}

In the previous analysis we found the emergence of power-law behavior near the horizon for specific NP quantities.  This behavior can be understood more geometrically through the notion of self-similarity in the near-horizon limit \cite{gralla2016near,Gralla:2017lto}.  We can assign an invariant weight to the self-similar tensor and use this notion to connect with other near-horizon phenomena, including the Meissner effect and the Aretakis instability. 

The near-horizon limit may be constructed by a coordinate transformation from BL coordinates $(t,r,\theta,\phi)$ to ``barred coordinates''  $x^{\bar{\mu}}=(\bar{t},\bar{r},\theta,\bar{\phi})$ defined by \cite{zaslavskii1998horizon,kunduri2007near,amsel2010uniqueness}
\begin{align}\label{scalingcoords}
    \bar{t} = \lambda \Omega_H t, \qquad \bar{r} = \frac{r-M}{\lambda M}, \qquad \bar{\phi} = \phi - \Omega_H t.
\end{align}
(In terms of the near-horizon deviation $x$, we have $\bar{r}=x/\lambda$.)

We denote barred coordinate components by placing a bar on the index, and place a bar on any tensor that is defined by a limiting process $\lambda \to 0$ in barred coordinates.  The KN near horizon geometry and Maxwell field are defined as
\begin{align}\label{gFweight}
    \bar{g}_{\bar{\mu} \bar{\nu}} = \lim_{\lambda \to 0} g_{\bar{\mu} \bar{\nu}}, \quad \bar{F}_{\bar{\mu} \bar{\nu}} = \lim_{\lambda \to 0} F_{\bar{\mu} \bar{\nu}}.
\end{align}
The details of the limiting values are not important; the crucial property is that the limiting metric remains non-degenerate (Lorentzian).  Since further rescalings do not change the limit, the resulting configuration is guaranteed to have an enhanced scaling symmetry $\bar{t} \to \bar{t}/C$ and $\bar{r} \to C \bar{r}$.\footnote{The near-horizon geometry also picks up an additional enhanced symmetry; this symmetry is ``accidental'' from the perspective of the scaling limit and not important for our considerations.}  This demonstrates the emergence of scaling symmetry near the KN horizon.

Perturbing fields will not, in general, possess the full scaling symmetry, but in many cases they have an emergent scaling \textit{self-similarity}, meaning that their barred coordinate components behave as a power law in $\lambda$ as $\lambda \to 0$.  We say that a tensor field $T_{\mu_1 \dots \mu_n}$ has \textit{weight} $w$ if\footnote{This is the opposite convention from our previous work  \cite{gralla2016near,Gralla:2017lto}, in which such a field would have weight $-w$.} 
\begin{align}\label{Tweight}
    T_{\bar{\mu}_1 \dots \bar{\mu}_n} \sim  \lambda^w \bar{T}_{\bar{\mu}_1 \dots \bar{\mu}_n}, \qquad \lambda \to 0,
\end{align}
where $\sim$ indicates asymptotic equality.  It follows from \eqref{Tweight} that the ``coefficient''\footnote{In general, when we speak of the weight of a tensor, we allow the possibility of a ``zero coefficient'' $\bar{T}_{\bar{\mu}_1 \dots \bar{\mu}_n}=0$.  For example, if we say that $T$ has weight one, its barred coordinates could in fact scale to zero faster than $\lambda$.  If the exact weight is known we will emphasize that fact.} $\bar{T}_{\bar{\mu}_1 \dots \bar{\mu}_n}$ is self-similar under further rescalings,
\begin{align}
   \bar{T} \to C^w \bar{T} \qquad \textrm{under} \quad   \bar{t}\to \bar{t}/C, \quad \bar{r} \to C \bar{r}.
\end{align} 
The weight of a tensor is additive under tensor product.  In particular, since the KN metric has weight 0 according to \eqref{gFweight}, the weight of a tensor is unaffected by raising/lowering indices.  In general, covariant operations involving the metric do not affect the weight of a tensor.  For example, the covariant derivative of a tensor has the same weight as the original tensor.

In our analysis of co-rotating perturbations, we found that Kinnersley-tetrad $\psi_s$ of a co-rotating mode behaves as $x^{w-s}$ (for some $w$) near the horizon,
\begin{align}\label{psisnearH}
    \psi_s\sim x^{w-s} S_s(\theta) e^{i m(\phi-\Omega_H t)} , \qquad x \to 0.
\end{align}
Changing to barred coordinates shows that
\begin{align}
    \psi_s\sim \lambda^{w-s} \bar{\psi}_s, \quad \lambda \to 0,
\end{align}
with
\begin{align}
    \bar{\psi}_s= \bar{r}^{w-s} S_s(\theta) e^{i m \bar{\phi}}.
\end{align}
Thus $\psi_s$ has weight $w-s$.  The co-rotating angular dependence is crucial for the existence of a well-defined weight; otherwise a $\lambda$ would appear in the exponential.

It is easy to check that the Kinnersley tetrad vectors have weights 
\begin{align}\label{Kinnersleyweight}
    \ell \sim \lambda^{-1} \bar{\ell}, \quad n \sim \lambda \bar{n}, \quad m\sim \lambda^0 \bar{m}.
\end{align}
Going back to the definition \eqref{psim1}--\eqref{psim2} and counting the appearances of $\ell$ and $n$ in each formula, one begins to suspect that the underlying tensors $\delta F_{\mu \nu}$ and $\delta C_{\mu \nu \rho \sigma}$ must have weight $w$ (at least ignoring trivial changes in mass, spin, and charge), such that the weights of the tetrad vectors account for the shifted weight $w-s$ of $\psi_s$.  

We can show this to be the case in the Kerr spacetime ($Q=0$) using the metric/field reconstruction procedure \cite{cohen1975space,chrzanowski1975vector,wald1973perturbations,wald1978construction}.  Schematically, the reconstructed metric may be written $h_{\mu \nu}=\mathcal{O}_{\mu \nu}[\psi_{-2}]$, where $\mathcal{O}_{\mu \nu}$ is a set of operations involving four angular integrals and then two derivatives.  (For details, we recommend the excellent presentation of Ref.~\cite{berens2024gravitational}.)  Using the Geroch-Held-Penrose \cite{GerochHeldPenrose1973, IulianoZahn2023} version of the latter step (e.g., Eq.~(19) of Ref.~\cite{Gralla:2017lto}), it is straightforward to establish that the reconstructed metric is invariant under tetrad boosts $\ell \to B \ell, n \to B^{-1}n$.  By using the boost $B=x$, we make the tetrad have weight $0$.  In this tetrad, $\tilde{\psi}_{-s}$ has weight $w$ and the metric reconstruction procedure manifestly preserves the weight (since it is built from the background metric and the weight-0 tetrad vectors), so that the resulting perturbation $h_{\mu \nu}$ also has weight $w$.  Wald's theorem \cite{wald1973perturbations}  guarantees that the reconstruction is unique up to gauge and changes in mass and spin.

Similar arguments establish that the Maxwell field $F_{\mu \nu}$ reconstructed from $\psi_{-1}$ in the Kerr spacetime has weight $w$.  We fully expect the same to be true of reconstructed metric and field perturbations in the KN spacetime, but we are unable to make an analogous argument at this time.

It is helpful to also introduce a version of this story better adapted to the regular coordinates and tetrad.  We instead consider ``tilde coordinates''  $x^{\tilde{\mu}}=(\tilde{v},\tilde{r},\theta,\tilde{\varphi})$
\begin{align}\label{tilde}
    \tilde{v} = \lambda \Omega_H v, \qquad \tilde{r} = \frac{r-M}{\lambda M}, \qquad \tilde{\varphi} = \varphi - \Omega_H v.
\end{align}
Near-horizon weights are defined in the same way, 
\begin{align}\label{Tweight2}
    T_{\tilde{\mu}_1 \dots \tilde{\mu}_n} \sim  \lambda^{\tilde{w}} \tilde{T}_{\tilde{\mu}_1 \dots \tilde{\mu}_n}, \qquad \lambda \to 0,
\end{align}
from which it follows that  
\begin{align}\label{limitingselfsim}
   \tilde{T} \to C^{\tilde{w}} \tilde{T} \qquad \textrm{under} \quad   \tilde{v}\to \tilde{v}/C, \quad \tilde{r} \to C \tilde{r}.
\end{align} 
Recalling \eqref{fKN}, a single co-rotating power-law may be expressed in either coordinate system as 
\begin{align}
    x^{w} e^{im(\phi-\Omega_H t)}  = x^{w+2iM\Omega_Hm} e^{im(\varphi-\Omega_H v)}.
\end{align}
The weights are thus related by 
\begin{align}\label{wwtilde}
    \tilde{w} = w + 2 i m M \Omega_H .
\end{align}
The weights are the same in the axisymmetric case $m=0$, and are otherwise related by a purely imaginary shift.  The weights $p$ defined in Ref.~\cite{Gralla:2017lto} are related to these weights by $p=-\tilde{w}$.

For future reference we note that the Hartle-Hawking tetrad vectors have weights
\begin{align}\label{HartleHawkingweight}
    \hat{\ell} \sim \lambda \hat{\bar{\ell}}, \quad \hat{n} \sim \lambda^{-1} \hat{\bar{n}}, \quad \hat{m}\sim \lambda^0 \hat{\bar{m}}.
\end{align}
Since the tetrad is axisymmetric, there is no distinction between $w$ and $\tilde{w}$.

\subsection{Comparison with other calculations}\label{sec:weight-comparison}

The invariant notion of tensor weight is useful for comparing results of different calculations.  We noted below \eqref{axih} that our axisymmetric numerical results are consistent with the analytic solution of Sec.~3 of Ref.~\cite{Horowitz:2024dch}.  To justify this claim, we recall that $w=h-1$ is the invariant weight of the regular solution with conformal weight $h$.  We found that axisymmetric solutions have conformal weight equal to an integer $h \geq 2$, meaning that regular axisymmetric solutions have invariant weight equal to an integer $w \geq 1$.  From Eqs.~(3.3)--(3.6) of Ref.~\cite{Horowitz:2024dch} we infer that the invariant weight of their analytic solution is $\gamma^{(0)}$.  Indeed, they find that $\gamma^{(0)}$ is an integer $\gamma^{(0)} \geq 1$, in agreement with our numerical calculations.\footnote{The additional family of solutions discussed in Sec.~3 of Ref.~\cite{Horowitz:2024dch} does not contain power-law behavior and hence is not relevant to our analysis.}  

We also noted below \eqref{RNh} that our findings for the RN limit $Q\to M$ are consistent with the analytic solution of Ref.~\cite{Porfyriadis:2018yag}.  To justify this claim, we again translate both results into statements about the invariant weight.  Our numerical analysis reveals two distinct families of conformal weights labeled by $h\to\ell$ for $\ell \geq 2$ and $h \to \ell+2$ for $\ell \geq 1$.  With each conformal weight is associated two distinct solutions with invariant weights $h-1$ and $-h$ (see Eq.~\eqref{wh}).  Thus we have the $\ell=1$ solutions with weights $\{2,-3\}$ together with four invariant-weight solutions $w=\{\ell-1,\ell+1,-\ell,-\ell-2\}$ for each $\ell \geq 2$.  Eq.~(27) of \cite{Porfyriadis:2018yag} also reveals four powers for each $\ell \geq 2$, which translate into invariant weights by adding two to each exponent (see (18) and (9) therein), producing the exact same list of four weights.  The $\ell=1$ case was not considered in Ref.~\cite{Porfyriadis:2018yag}.

Our work is also consistent with the near-extremal quasinormal mode study of Dias, Godazgar, and Santos~\cite{DiasGodazgarSantos2022}.  These authors managed to separate the perturbation equations in the near-horizon limit, and obtained an analytic solution for the radial dependence in terms of a separation constant  $\lambda_2$.  This solution has power-law exponents which, when expanded in the overlap region (their equation (3.38)), correspond to our power-law exponents after the identification
\begin{equation}
 h=\frac{1}{2}+\frac{\sqrt{\lambda_{2}}}
 {2(1+\hat a^2)}, \quad \hat{a}\equiv \frac{a}{M}.
\end{equation}

Ref.~\cite{DiasGodazgarSantos2022} solved for $\lambda_2$ numerically by solving the coupled angular equations (similar to our approach) and also found an analytic solution in the large-$m$ limit (see (3.33) therein),
\begin{equation}
 \lambda_2=4(1-4\hat{a}^2)m^2+O(m).
\end{equation}
The threshold between principal and supplementary is $\lambda_2=0$, meaning $\hat a=1/2$ and hence $Q/M=\sqrt{3}/2$.  This large-$m$ threshold was observed numerically in Fig.~\ref{fig:Qc_vs_m}.

The same number $\sqrt{3}/2$ appears as the threshold for the existence of timelike co-rotating orbits outside the horizon.   The similarity may be seen mathematically by expanding the norm of the horizon killing field $\chi=\partial_t+\Omega_H\partial_\phi$ on the equator near the horizon,
\begin{equation}
 \chi^2
 =\frac{4\hat a^2-1}{(1+\hat a^2)^2}x^2+O(x^3).
 \label{eq:chi-equatorial-expansion}
\end{equation}
For $\hat{a} \neq 1/2$ it follows immediately that 
\begin{equation}
 h
 =\frac{1}{2} + \sqrt{-m^2\lim_{x\to0}\frac{\chi^2}{x^2}}+O(1).
 \label{eq:h-chi-relation}
\end{equation}
Thus a spacelike $\chi$ just outside the horizon gives a principal weight
$h=1/2+i\delta$, whereas a timelike $\chi$ gives a real supplementary
weight.

\section{Meissner effect}\label{sec:Meissner}

In the axisymmetric case $m=0$, our co-rotating perturbations are stationary.  That is, $m=0$ is the simpler case of stationary, axisymmetric perturbations.  For EM perturbations of the Kerr spacetime, the EM scaling weights are integers and give rise to the Meissner effect \cite{gralla2016near}.  We now review this result and generalize it to GEM perturbations of the extremal KN spacetime.

For an extremal Kerr black hole with stationary, axisymmetric perturbations, the general solution regular at the poles is
\begin{align}\label{psisum}
    \psi_s = \sum_{\ell=|s|}^\infty\left( a_\ell x^{\ell-s} + b_\ell x^{-(\ell+1)-s}\right)P^{(s)}_{\ell}(\cos \theta),
\end{align}
where $P^{(s)}_{\ell}$ are associated Legendre Polynomials.  The $b_{\ell}$ family is singular on the horizon and discarded.  From the $a_{\ell}$ family we immediately conclude that for regular perturbations, each $\ell$-mode of $\psi_{s}$ has weight $\ell-s$.  Below Eq.~\eqref{Kinnersleyweight} we established that the reconstructed perturbation (maxwell field or metric perturbation) has weight $\ell$, up to gauge and trivial changes in mass, charge and spin.

Restricting to stationary, axisymmetric electromagnetic perturbations of extremal Kerr that do not change the charge of the spacetime, the general solution is thus
\begin{align}\label{F}
    F_{\mu \nu} = \sum_{\ell=1}^\infty F^{(\ell)}_{\mu \nu}(r,\theta), \quad F^{(\ell)}_{\tilde{\mu}\tilde{\nu}} \sim \lambda^\ell.
\end{align}
where $F^{(\ell)}_{\mu \nu}$ has weight $\ell$ as indicated by $F^{(\ell)}_{\tilde{\mu}\tilde{\nu}} \sim \lambda^\ell$.  Here we use the ingoing version of the near-horizon limit since we will be evaluating tensors on the future horizon.  The weight $\tilde{w}=w=\ell$ is the same as the BL version for stationary, axisymmetric perturbations.

To connect to the traditional Meissner screening, consider expressing this field in co-rotating regular coordinates $(v,r,\theta,\tilde{\varphi})$.  Changing to the tilde coordinates \eqref{tilde} entails $v \to \tilde{v}/(\Omega_H\lambda)$ and $r \to M(1+\lambda \tilde{r})$, so that expanding a $v$-independent (stationary) tensor component in $\lambda$ is just a Taylor expansion in radius about the horizon, together with an overall weighting depending on how many times $v$ and $r$ appear in the component.  For example,
\begin{align}
    F_{\tilde{v}\tilde{\theta}} = \frac{1}{\Omega_H\lambda}\!\Big( & F_{v\theta} + \lambda M\partial_r F_{v\theta} \tilde{r} \nonumber \\ & + \frac{1}{2}M^2\lambda^2 \partial_r \partial_r F_{v\theta} \tilde{r}^2 + ... \Big)_{\!\!r=M}\!.
\end{align}
The general form \eqref{F} requires that this perturbation vanish until order $\lambda^1$ (the lowest weight-term being $\ell=1$), which means that $F_{v\theta}$ and its first derivative must vanish on the horizon, i.e. $F_{v\theta} = O((r-M)^2)$.  Making a similar argument for the other components, the complete collection behaves as
\begin{align}
F_{v\theta} \sim F_{v\tilde{\varphi}} \sim (r-M)^2&\ \label{Fscale1} \\
F_{v r} \sim F_{\theta \tilde{\varphi}} \sim  (r-M)^1&\ \label{Fscale2} \\
F_{r\theta} \sim F_{r\tilde{\varphi}} \sim (r-M)^0&.\ \label{Fscale3} 
\end{align}
In particular, the horizon pullback of $F$ and its dual are vanishing, which is the invariant statement of the electromagnetic black hole Meissner effect.

Turning now to gravitational perturbations of the extremal Kerr spacetime, we have established that the general regular, stationary, axisymmetric perturbation, which does not change the mass or spin of the spacetime, may be written as
\begin{align}\label{h}
    h_{\mu \nu} = \sum_{\ell=2}^\infty h^{(\ell)}_{\mu \nu}(r,\theta) + \textrm{gauge}, \qquad h^{(\ell)}_{\tilde{\mu}\tilde{\nu}} \sim \lambda^\ell,
\end{align}
where $h^{(\ell)}_{\mu \nu}$ has weight $\ell$, as indicated by $h^{(\ell)}_{\tilde{\mu}\tilde{\nu}} \sim \lambda^\ell$.

Again transforming to co-rotating coordinates $(v,r,\theta,\tilde{\varphi})$ and using the requirement from \eqref{h} that this perturbation vanish until order $\lambda^2$ (the lowest weight-term being $\ell=2$), we find (up to gauge) that
\begin{align}
h_{vv} \sim (r-M)^4&\ \\
h_{v\theta} \sim h_{v\tilde{\varphi}} \sim (r-M)^3&\ \\
h_{v r} \sim h_{\theta \theta} \sim h_{\theta \tilde{\varphi}} \sim h_{\tilde{\varphi} \tilde{\varphi}} \sim (r-M)^2&\ \\
h_{r\theta} \sim h_{r\tilde{\varphi}} \sim (r-M)^1&\  \\
h_{rr} \sim (r-M)^0&.
\end{align}
In particular, only $h_{rr}$ is nonvanishing on the horizon.  The  vanishing of the collection $\{h_{vv},h_{v\theta},h_{v\tilde{\varphi}},h_{\theta \theta}, h_{\theta \tilde{\varphi}}, h_{\tilde{\varphi} \tilde{\varphi}}\}$ is equivalent to the  invariant statement that the horizon pullback of the metric perturbation is pure gauge.  In this sense, the gravitational perturbation does not penetrate the horizon: a gravitational Meissner effect.

The existence of a gravitational Meissner effect suggests that the gravitational decoherence induced by black holes \cite{danielson2022black,danielson2023killing,gralla2024decoherence,biggs2024comparing} may disappear in the extremal limit, as it does for photon-mediated decoherence \cite{gralla2024decoherence}.  (However, see Ref.~\cite{Biggs:2026zlp} for an alternative point of view.)  This vanishing-decoherence result holds for a stationary observer placed on the black hole symmetry axis.  If one instead considers a co-rotating observer, then our results suggest \textit{amplified} decoherence in the extremal limit.

Finally we turn to GEM perturbations of the KN spacetime.  For axisymmetric perturbations, we have found positive-integer invariant weights $w \geq 1$, in agreement with analytic solutions \cite{Horowitz:2024dch}.  As long as the reconstructed fields still share the symmetries and invariant weights (see discussion near the end of Sec.~\ref{sec:weight}), we will have the same scalings \eqref{Fscale1}--\eqref{Fscale3} for the electromagnetic field, and the modified scalings for the reconstructed metric perturbation, 
\begin{align}
h_{vv} \sim (r-M)^3&\ \\
h_{v\theta} \sim h_{v\tilde{\varphi}} \sim (r-M)^2&\ \\
h_{v r} \sim h_{\theta \theta} \sim h_{\theta \tilde{\varphi}} \sim h_{\tilde{\varphi} \tilde{\varphi}} \sim (r-M)^1&\ \\
h_{r\theta} \sim h_{r\tilde{\varphi}} \sim h_{rr} \sim (r-M)^0&.
\end{align}
(The last line simply notes that the weight of the perturbation imposes no restriction on these components.)  In particular, the pullback of the metric to the future horizon $r=M$ is still vanishing.  Thus we may state the GEM Meissner effect as follows: 

\noindent \textit{Regular, stationary, axisymmetric  perturbations $\{h_{\mu \nu},\delta F_{\mu \nu}\}$ of the extremal KN black hole that leave its mass, spin, and charge invariant have vanishing (pure-gauge) pullback to the future horizon.}

Although we are confident in the correctness of this claim, our derivation does rely on unproven assumptions about the field reconstruction procedure for KN spacetime---see discussion at the conclusion of Sec.~\ref{sec:weight}.

The Meissner screening does not extend to stationary, \textit{nonaxisymmetric} fields.\footnote{In the special case $Q=M$ where the background is spherically symmetric (RN spacetime), an arbitrary stationary perturbation can be expressed as a sum over individual modes each axisymmetric about some axis, and hence the Meissner effect does occur even for nonaxisymmetric perturbations.}  To see why, consider expressing such a field (taken scalar for simplicity) in scaling coordinates \eqref{scalingcoords},
\begin{align}
    \psi & =\psi(r,\theta,\varphi) \\ 
    & = \psi\left(M(1+\lambda \tilde{r}), \theta,\tilde{\varphi}+\frac{\tilde{v}}{\lambda}\right).
\end{align}
When $\psi$ is axisymmetric, there is no dependence on the last slot, and expanding in $\lambda$ recovers the near-horizon power-law behavior in $r-M$ discussed previously.  However, for non-axisymmetric $\psi$, the limit $\lambda \to 0$ is clearly singular, wrapping around the circle an infinite number of times.  Therefore we see that regular, stationary, nonaxisymmetric perturbations \textit{can never be self-similar under near-horizon scaling}.  This provides a simple explanation for why the Meissner screening fails.

\section{Aretakis instability}\label{sec:Aretakis}

The Aretakis instability can be understood as the emergence of self-similarity at late times near the horizon \cite{Gralla:2017lto}.  The near-horizon limit fixes $x/\lambda$ and $\lambda v$ as $\lambda \to 0$ and hence flows simultaneously to the horizon and to late times (provided $v>0$).  For a scalar field $\psi$, weight-$\tilde{w}$ self-similarity requires the limiting field to obey \eqref{limitingselfsim}, fixing its form to
\begin{align}
    \tilde{\psi} = \tilde{v}^{-\tilde{w}} f(\tilde{r}\tilde{v},\theta,\tilde{\varphi}),
\end{align}
with a smooth function $f$ guaranteeing horizon-regularity.  The original field \eqref{Tweight2} then obeys
\begin{align}
    \psi & = \lambda^{\tilde{w}}(\lambda \Omega_H v)^{-\tilde{w}} f\left(\frac{(r-M)\Omega_H}{M}v,\theta,\tilde{\varphi}\right)\\
    &= v^{-\tilde{w}} g((r-M)v,\theta,\tilde{\varphi}),
\end{align}
in the limit $r\to M$, $v \to \infty$, with $(r-M)v$ fixed.  In the second line we introduced a new smooth function $g$ for convenience.  We may use this form to evaluate the $n{}^{\rm th}$ radial derivative on the horizon,
\begin{align}
    \left.(\partial_r)^n\psi\right|_{r=M} = v^{-\tilde{w}+n} g^{(n)}(0,\theta,\tilde{\varphi}),
\end{align}
where the superscript $(n)$ refers to $n$ derivatives with respect to the first slot of $g$.  As long as the real part of the weight $\tilde{w}$ is positive, the field will decay.  However, we see the characteristic Aretakis behavior that successive radial derivatives decay one power slower, and eventually grow.

Ref.~\cite{Gralla:2017lto} showed that freely decaying perturbations of the Kerr spacetime may be written at late times as an infinite sum of definite-weight terms,\footnote{The convergence of the sum was not analyzed.} confirming this explanation as the origin of Aretakis behavior.  For non-axisymmetric modes, the real parts of the calculated weights agree with the real parts of the conformal weight $h$ (using our sign conventions, and noting that the difference \eqref{wwtilde} between $w$ and $\tilde{w}$ is a shift of imaginary part).  Although this relationship has not been checked for the KN spacetime, the underlying mathematical similarity  strongly suggests that it will continue to hold at non-zero black hole charge.  We therefore predict that at late times near the horizon, generic GEM perturbations of the extremal KN spacetime will take the form of an infinite sum of definite-weight perturbations with weight satisfying\footnote{We only discuss agreement of the real parts because of the imaginary phase shift \eqref{wwtilde} relating $w$ and $\tilde{w}$ and because of the additional phase shifts present for late-time weights (the parameter $n$ in Eq.~(15) of \cite{Gralla:2017lto}, in which the notation $p=-\tilde{w}$ was used). It is worth emphasizing that the single conformal weight $h$ controls two distinct phenomena: co-rotating perturbations have weight with $\textrm{Re}[w]=\textrm{Re}[h-1]$ meaning $\textrm{Re}[w]=-1/2$ for principal modes, while freely decaying late-time perturbations have weight $\textrm{Re}[w]=\textrm{Re}[h]$, meaning $\textrm{Re}[w]=+1/2$ for principal modes.  }
\begin{align}
    \textrm{Re}[w] = \textrm{Re}[h],
\end{align}
where $h$ are the conformal weights calculated numerically in this paper.  Higher values of $\textrm{Re}[w]$ decay faster, so the lowest value ``wins'', predicting the late-time behavior.  The axisymmetric modes will decay faster and can be neglected.\footnote{For axisymmetric modes, Ref.~\cite{Gralla:2017lto} found conformal weight $h=\ell+1$ and late-time weight $w=\ell+2$ (written in our conventions). The KN axisymmetric conformal weights we compute are also $h=2,3,4,\dots$, so we expect late-time axisymmetric weights of $w=3,4,5,\dots$.  From Fig.~\ref{fig:minh} we see that there always exists a non-axisymmetric conformal weight with $\textrm{Re}[h] \leq 2$, implying a late-time invariant weight $w \leq 2$.  Thus the late-time axisymmetric weights $w \geq 3$ can always be neglected.}

The precise prediction for any given quantity can be made with the help of Tab.~\ref{tab:kn-physical-h-grid} and the rules for manipulating the tensor weight.  For example, consider a KN black hole with charge $Q/M=0.94$.  From Tab.~\ref{tab:kn-physical-h-grid} we see that there are no principal modes, with the smallest $\textrm{Re}[h]$ being the supplementary mode $h\approx0.9336$ at $m=2$.  This mode will win at late times, and the GEM perturbations will have weight $w \approx 0.9336$ with $m=2$ azimuthal dependence.  The growth or decay of a scalar can then be constructed by adding up the weights of its constituent tensors.  For example, $\hat\Psi_4\propto\hat{\psi}_{-2}$ is built from the weight-$w$ perturbation and two copies of the Hartle-Hawking tetrad $\hat{n}$, which has weight equal to $-1$ (see \eqref{HartleHawkingweight}).  The weight is thus $w-2$, so $\hat\Psi_4$ will behave as $v^{-w+2}$, i.e., it will grow like $v^{1.066}$.  Similarly, any principal mode will grow like $v^{3/2}$, as in the Kerr spacetime.   At least when there are finitely many principal modes \eqref{finite} (so the mode sum will not introduce worse behavior), a generic perturbation will exhibit this behavior. 

We also expect off-horizon power-law decay with a rate precisely twice that of the on-horizon decay, without any enhancement from derivatives or scaling weights \cite{Gralla:2018xzo}.  In particular, all Weyl scalars should decay like $t^{-2w}$, where $w$ is the smallest $\textrm{Re}[h]$.  We thus predict $t^{-1}$ decay when there are principal modes (at least when there are a finite number) and faster decay otherwise.  For the example $Q=.94M$ of the prior paragraph, we predict $t^{-1.867}$ decay, dominated by the $m=2$ mode. It would be very interesting to check these predictions with numerical studies \cite{Zilhao:2014cfa,mark2015quasinormal,Dias:2015wqa,Dias:2021yju} or mathematically rigorous arguments  \cite{Angelopoulos:2018uwb,Gajic:2023uwh,Gajic:2026aah,Gajic:2026mre,Fang:2026oyg}.

It is useful to directly compare the Aretakis instability with the amplification effect studied here.  Both phenomena involve co-rotating modes and are controlled by the same conformal weight $h$, so there is clearly an intimate relationship.  The Aretakis instability involves growth of curvature on the horizon (e.g., $\hat{\Psi}_4\sim v^{3/2}$ in Kerr).  For near-extremal black holes of surface gravity $\kappa$, the growth occurs over a coordinate distance scaling as $x\sim\kappa$ and timescale $\Delta v \sim 1/\kappa$ \cite{Gralla:2016sxp}.  Infalling observers experience this curvature of up to $\sim \kappa^{-3/2}$ over a proper time $\sim \kappa$, and integrating twice predicts a largest possible tidal deformation scaling as $\kappa^2 \kappa^{-3/2}\sim\sqrt{\kappa} \to 0$.  Thus the Aretakis instability involves large curvatures but small tidal deformation.

By contrast, the amplification studied in this paper involves formally-infinite tidal deformation.  The sources we consider have a non-zero co-rotating harmonic and are in effect continuously driving the system at the frequency where the Aretakis behavior occurs.  It is therefore natural to view the horizon amplification as a driven Aretakis instability.  It is notable that the driving source not only promotes the growing curvatures of the Aretakis instability to formally-infinite curvatures, but also promotes the \textit{decaying} tidal deformation of the Aretakis instability to formally-infinite tidal deformation.  Indeed, the linear approximation appears to break down entirely in the presence of co-rotating sources.

\section*{Acknowledgements}
We are grateful to Dejan Gajic, Gary Horowitz, and Alex Lupsasca for helpful conversations, and to O. Dias and J. Santos for additionally sharing unpublished notes.   This work was supported by grants from the Simons Foundation (MPSCMPS-00001470) and the National Science Foundation (PHY-230919).

\appendix

\section{A co-rotating Vlasov source from Kerr geodesics}
\label{app:resonant-vlasov-source}

In this appendix we give an explicit construction of a conserved, positive-energy,
non-axisymmetric source which is invariant under the horizon-generating Killing
field of extremal Kerr.  The source is collisionless matter, described by a
Vlasov distribution on the future timelike mass shell.  The construction relies on the existence of
coordinate-time action-angle variables for regular bound timelike geodesics
\cite{Carter:1968rr,Schmidt:2002qk,Drasco:2003ky}.

We write the Kerr-geodesic action-angle variables as
\begin{align}
(J_A,w^A),\qquad A\in\{r,\theta,\phi\},
\end{align}
so that the geodesic flow is
\begin{align}
\frac{dJ_A}{dt}=0,
\qquad
\frac{dw^A}{dt}=\Omega_A(J),
\end{align}
with $\Omega_A$ the associated frequencies.  
The symbol $J$ denotes the collection $\{ J_r,J_\theta,J_\phi\}$.  From the explicit form
\begin{align}
   W=J_\phi \phi+ A(r;J) + B(\theta;J) 
\end{align}
of the generating function of the canonical transformation, it follows that all the actions and angles are independent of $\phi$ and $t$, except for the angle $w^\phi$ which (noting $J_\phi=p_\phi$) satisfies
\begin{align}
    w^\phi - \phi=\textrm{(independent of $\phi$ and $t$)}.
\end{align}
When expressed in angle-action coordinates, a co-rotating (i.e. helically symmetric) scalar field is therefore a function of $w^\phi-\Omega_H t$.

A simple way to solve the Vlasov equation is to take the distribution function $f$ to depend only on the conserved actions, so that $df/dt=0$ automatically.  However, this produces a stationary, axisymmetric distribution function.  To create a co-rotating distribution, we want to allow dependence on $w^\phi-\Omega_H t$, while preventing this dependence from spoiling the Vlasov property.  This may be done by introducing a ``phase field''
\begin{align}
\Psi_k
=
k_r w^r+k_\theta w^\theta
+m\bigl(w^\phi-\Omega_H t\bigr)
\end{align}
for each choice of integers $
k_r,k_\theta,m$.  Our strategy will be to include dependence on this phase, while also including a delta function of its time derivative $\mathcal{R}_k=d\Psi/dt$, 
\begin{align}\label{eq:resonance-cond}
\mathcal{R}_k =
k_r \Omega^r+k_\theta \Omega^\theta
+m\bigl(\Omega_\phi-\Omega_H \bigr).
\end{align}
Given functions $F$ and $G$, we construct a Vlasov distribution by
\begin{align}
f
= \Theta(p^t) \delta(g_{\mu \nu}p^\mu p^\nu+\mu^2)
\,
\delta(\mathcal R_k(J))G(\Psi_k)F(J).
\end{align}
The first two factors select the positive mass shell.  The third factor $\delta(\mathcal{R}_k)$ restricts to $\mathcal{R}_k=0$, where $d\Psi_k/dt=0$ by construction.  Since $J=\{J_r,J_\theta,J_\phi\}$ and $g_{\mu \nu}p^\mu p^\nu$ are also conserved quantities, this provides a valid Vlasov solution $df/dt=0$ as long as the formula is well-defined and non-zero.  It is helically symmetric by construction.

We may ensure a well-defined, non-zero solution by meeting the following conditions.  First, one must choose the support of $F$ to be entirely within the region of bound orbits, where $(J_A,w^A)$ can be determined from $x^\mu$ and $p^\mu$.  One should also ensure non-zero intersection with the set $\mathcal{R}_k=0$ for the given choice of $k=\{k_r,k_\theta,m\}$, as well as the regularity condition $\nabla_{J_A}\mathcal{R}_k\neq0$.  To represent a distribution function, $F$ and $G$ must be non-negative, and $G(\Psi_k)$ should be $2\pi$-periodic, in order to respect the identification $w^A\sim w^A+2\pi$ of the angle variables it relies on.  For a non-axisymmetric solution, we further want $G$ non-constant and $m\neq 0$. 

As a definite example, we may choose $k=(3,0,1)$ and take the support of $F$ to lie in a small neighborhood of the orbit discussed in Sec.~\ref{sec:Kerr} (see \eqref{kerrRes}).  Then the choice $G=1+\cos(2\Psi)/2$ excites the $m=2$ principal mode.

\section{Numerical calculation of the near-horizon exponents}\label{app:numerical_exp}

In this appendix we explain the numerical methods underlying the near-horizon exponents and eigenfunctions presented in Sec.~\ref{sec:KN-exponents}.  We set $\omega=m \Omega_H$ (i.e., $k=0$) and make the ansatz
\begin{align}
\psi_{-1}(x,\theta)=&x^{p}S_{-1}(\theta),\\
\psi_{-2}(x,\theta)=&x^{p+1}S_{-2}(\theta),
\end{align}
in the perturbation equations \eqref{eq:PDE1}--\eqref{eq:PDE2}.  Keeping only the leading term in $x$ near zero, we obtain a pair of coupled second-order ordinary differential equations on the sphere,
\begin{align}\label{angularode1app}
\mathcal{L}_{-2}[S_{-2}(\theta),S_{-1}(\theta)] &= 0,\\
\mathcal{L}_{-1}[S_{-2}(\theta),S_{-1}(\theta)] &= 0.\label{angularode2app}
\end{align} 
The coefficients in these linear differential operators depend smoothly on $\theta$, $m$, and $Q$, and polynomially on $p$ up to order $p^2$.  The ansatz presented in the main text is equivalent to the one used here after identifying
\begin{align}
    w=p-1.
\end{align}

Regularity at the poles fixes the allowed local behavior of the angular functions. Expanding the coupled system near $\theta=0$ and $\theta=\pi$ and discarding the singular solution gives
\begin{align}
S_s(\theta)&\sim \theta^{|m+s|}\qquad\qquad\ \  (\theta\to0),
\\
S_s(\theta)&\sim (\pi-\theta)^{|m-s|}\qquad (\theta\to\pi),
\end{align}
for spin weights $s=-1$ and $s=-2$. If the corresponding pole exponent is nonzero, the field must vanish there, so we impose a Dirichlet condition. If the exponent vanishes, the field is finite and generically nonzero, and regularity instead requires a Neumann condition. Thus at each pole we impose
\begin{align}
S_s(0)=0
&\quad \text{if} \quad |m+s|>0,\\
S_s(\pi)=0
&\quad \text{if} \quad |m-s|>0,\\
\partial_\theta S_s(0)=0
&\quad \text{if} \quad |m+s|=0,
\\
\partial_\theta S_s(\pi)=0
&\quad \text{if} \quad |m-s|=0.
\end{align}

To discretize the problem we use Chebyshev--Lobatto collocation
\cite{Trefethen2000Spectral,WeidemanReddy2000DiffSuite}. We begin with the standard Chebyshev nodes
\begin{equation}
y_j=-\cos\!\left(\frac{j\pi}{N}\right),\qquad j=0,\dots,N,
\end{equation}
on $[-1,1]$, and map them linearly to the angular interval by
\begin{equation}
\theta_j=\frac{\pi}{2}(y_j+1).
\end{equation}
Given the $N+1$ nodal values of some function $f(\theta)$,
\begin{align}
f_j=f(\theta_j),
\end{align} 
the Chebyshev interpolating polynomial
is the unique degree-$N$ polynomial in the variable
\begin{equation}
y=\frac{2\theta}{\pi}-1
\end{equation}
that agrees with $f$ at the nodes.  The interpolation $\mathcal{I}_N f$ is written in the cardinal basis
as
\begin{equation}
(\mathcal I_N f)(\theta)
=
\sum_{j=0}^{N} f_j \ell_j(\theta),
\end{equation}
where the cardinal functions are
\begin{equation}\label{def-cardinal}
\ell_j(\theta)
=
\prod_{\substack{k=0\\ k\neq j}}^{N}
\frac{y(\theta)-y_k}{y_j-y_k},
\qquad
\ell_j(\theta_i)=\delta_{ij}.
\end{equation}
The first-derivative matrix is defined by differentiating these cardinal
functions and evaluating again at the nodes,
\begin{equation}
(D_\theta)_{ij}=\ell_j'(\theta_i).
\end{equation}
The diagonal entries are fixed by the requirement that the derivative of a
constant vanish:
\begin{equation}
(D_\theta)_{ii}=-\sum_{\substack{j=0\\ j\neq i}}^{N}(D_\theta)_{ij}.
\end{equation}
Thus, for the nodal vector
\begin{equation}\label{f_vector}
    f=(f_0,\dots,f_N)^T,
\end{equation} 
$D_\theta f$ is the vector of
nodal values of $\partial_\theta(\mathcal{I}_N f)$.  That is, the first-derivative matrix $D_\theta$ acts on nodal values and returns the derivative of the Chebyshev interpolating polynomial at the same nodes.  Similarly, 
$D_\theta^2 f$ gives the nodal values of
$\partial_\theta^2(\mathcal{I}_N f)$. Differentiation has therefore been
converted into matrix multiplication. 

The pole conditions are imposed by a lift from interior values to full nodal values,
equivalently eliminating the endpoint degrees of freedom in the collocation
system \cite{Trefethen2000Spectral,WangSamsonZhao2014PSIM}.
For each spin sector $s$, we define the two boundary rows
\begin{align}
B_s^{(0)}
&=
\begin{cases}
e_0^T, & |m+s|>0,\\
e_0^T D_\theta, & |m+s|=0,
\end{cases}
\\
B_s^{(\pi)}
&=
\begin{cases}
e_N^T, & |m-s|>0,\\
e_N^T D_\theta, & |m-s|=0,
\end{cases}
\end{align}
where $e_j$ is the column vector selecting the $j$-th node (components $e^i_j=\delta^i{}_j$), and a $T$ denotes transpose.  We combine these into a $2\times(N+1)$ matrix as
\begin{equation}
B_s=
\begin{pmatrix}
B_s^{(0)}\\
B_s^{(\pi)}
\end{pmatrix}.
\end{equation}
From the definition \eqref{f_vector}, the discrete regularity condition is then simply
\begin{equation}\label{liftbc}
B_s f=0.
\end{equation}

The boundary condition \eqref{liftbc} allows a unique association between full nodal vectors $f$ and ``interior vectors'' $\widehat{f}$ defined only on the interior nodes $j=\{1,\dots,N-1\}$.  Restriction from full to interior can be written
\begin{align}
    \widehat{f}=Rf,
\end{align}
where $R$ is the $(N-1)\times(N+1)$ matrix that preserves the interior values of $f$.  For the other direction, we define the lift $P_s$ (for each $s$) as the unique $(N+1)\times(N-1)$ matrix satisfying
\begin{equation}
R P_s=\mathbb{1},
\qquad
B_s P_s=0.
\end{equation}
An interior vector $\widehat{f}$ is then uniquely lifted to a full vector satisfying the boundary conditions \eqref{liftbc} by
\begin{align}
    f=P_s\widehat f.
\end{align}

We will work with the $2(N+1)$-dimensional vector containing the nodal values of $S_{-1}$ and $S_{-2}$,
\begin{equation}
 U=
\begin{pmatrix}
 u\\
 v
\end{pmatrix}, \quad 
u_j=S_{-1}(\theta_j),\quad v_j=S_{-2}(\theta_j).
\end{equation}
Similarly, we define a hatted version living on the interior grid only,
\begin{equation}\label{def:p-matrix}
\widehat U=
\begin{pmatrix}
\widehat u\\
\widehat v
\end{pmatrix},
\qquad
\mathcal P=
\begin{pmatrix}
P_{-1} & 0\\
0 & P_{-2}
\end{pmatrix},
\qquad
U=\mathcal P\widehat U.
\end{equation}

Substituting the Chebyshev interpolants for $S_{-1}$, $S_{-2}$, and their
derivatives into the coupled angular equations \eqref{angularode1app}--\eqref{angularode2app} gives a matrix equation of the form
\begin{equation}
(\mathcal M_0+p\mathcal M_1+p^2\mathcal M_2)U=0,
\end{equation}
where $\mathcal{M}_i$ are $2(N+1)$-dimensional square matrices that depend on $m$ and $Q$ (independent of $p$).

We enforce the differential equations only at the interior nodes and impose the endpoint equations through the lift. Introducing 
\begin{equation}
\mathcal R=
\begin{pmatrix}
R & 0\\
0 & R
\end{pmatrix},
\end{equation}
the interior matrices are 
\begin{equation}
\widehat{M}_i=\mathcal R\,\mathcal M_i\,\mathcal P,
\qquad
i=0,1,2.
\end{equation}
The spectral problem is therefore the generalized quadratic eigenvalue problem
\begin{equation}
\bigl(\widehat{M}_0+p\widehat{M}_1+p^2\widehat{M}_2\bigr)\widehat U=0.
\end{equation}
After an eigenvector $\widehat U$ is found, the corresponding full angular
eigenfunctions are recovered by the lift $U=\mathcal P\widehat U$.

We solve the quadratic problem by companion linearization
\cite{TisseurMeerbergen2001QEP}. Introducing the augmented vector
\begin{equation}
W=
\begin{pmatrix}
\widehat{U}\\ p\widehat{U}
\end{pmatrix},
\end{equation}
the equation is rewritten as a generalized linear eigenvalue problem
\begin{equation}
A W=p\,B W,
\end{equation}
where
\begin{equation}
A=
\begin{pmatrix}
0 & \widehat{\mathbb{1}}\\
-\widehat{M}_0 & -\widehat{M}_1
\end{pmatrix},
\qquad
 B=
\begin{pmatrix}
\widehat{\mathbb{1}} & 0\\
0 & \widehat{M}_2
\end{pmatrix},
\end{equation}
where $\widehat{\mathbb{1}}$ is the $2(N-1)$-dimensional identity matrix. 
We solve this generalized eigenvalue problem using MATLAB's \texttt{eig} routine \cite{MATLAB}. The output gives the candidate exponents $p$, together with the corresponding interior eigenvectors $\widehat U$. We retain the modes that remain stable as the spectral resolution $N$ is increased. Physical eigenvalues exhibit the expected exponential convergence with $N$, while spurious modes drift under changes of resolution and are discarded.

As a check, in the Kerr limit $Q\to0$ the convergent eigenvalues reproduce the Kerr exponents \eqref{Kerrh}.  We also compared the
corresponding numerical eigenfunctions against spin-weighted
spheroidal harmonics computed independently with the Python package
\texttt{qnm}~\cite{Stein:2019mop}.  The checked modes agree in $h$ at
the $10^{-9}$ level or better, and the normalized angular-profile
overlaps are unity to numerical precision.

\section{Numerical construction of the Green's function}\label{app:green}

In this appendix we describe the numerical method used to calculate the retarded Green function near co-rotation (Sec.~\ref{sec:KNgreens} above).

For fixed azimuthal number $m$ and frequency detuning $k$, we write the coupled fields as
\begin{equation}\label{def:psi}
\Psi(r,\theta)
=
\begin{pmatrix}
\psi_{-1}(r,\theta)\\
\psi_{-2}(r,\theta)
\end{pmatrix}.
\end{equation}
The frequency-domain Kerr--Newman perturbation equations may be written as
\begin{equation}
\mathcal L(k)\Psi(r,\theta)
=
\mathcal J(r,\theta),
\qquad
\end{equation}
where the operators and fields are matrix-valued. The operator
$\mathcal L(k)$ is the coupled radial-angular operator obtained
from the Chandrasekhar--Kerr--Newman equations after fixing $(\omega,m)$.  We will generally write the $k$-dependence of this operator explicitly, while suppressing its $m$-dependence.

The corresponding retarded Green function is the matrix kernel
\begin{equation}
\mathcal{G}(r,\theta;r_s,\theta_s;k),
\end{equation}
defined by 
\begin{equation}
\mathcal L(k)
\mathcal{G}(r,\theta;r_s,\theta_s;k)
=
\mathbb{1}\,
\delta(r-r_s)\,\delta(\theta-\theta_s),
\end{equation}
together with retarded radial boundary conditions. Here $\mathbb{1}$ is the two dimensional identity matrix of the spin coefficients.  This is a different normalization than we used in the Kerr case \eqref{radialeqnSource}.  With this convention, a source produces the field 
\begin{equation}
\Psi(r,\theta)
=
\int dr_s\int_0^\pi d\theta_s
\mathcal{G}(r,\theta;r_s,\theta_s;k)
\mathcal J(r_s,\theta_s).
\end{equation}

We now discretize in the angular direction, using the  Chebyshev collocation grid and pole-regular lift described in Appendix \ref{app:numerical_exp}.   We introduce
\begin{align}
n=2(N-1)
\end{align}
for convenience.  The continuum operator $\mathcal{L}$ becomes a matrix-valued radial derivative operator $L$, 
\begin{equation}\label{Lk}
L(k)
=
A_2(r,k)\frac{d^2}{dr^2}
+
A_1(r,k)\frac{d}{dr}
+
A_0(r,k),
\end{equation}
where $A_i$ are $n\times n$ matrices.

For source normalization we use Fejér's second quadrature rule on the same
Chebyshev--Lobatto grid~\cite{Waldvogel2006}. Equivalently, this is the interpolatory quadrature
rule obtained from the interior nodes $\theta_1,\ldots,\theta_{N-1}$. Let $\lambda_j(\theta)$ denote the
corresponding interior cardinal functions,
\begin{equation}
\lambda_j(\theta)
=
\prod_{\substack{k=1 \\ k\ne j}}^{N-1}
\frac{y(\theta)-y_k}{y_j-y_k},
\qquad
\lambda_j(\theta_i)=\delta_{ij},
\end{equation}
with $i,j=1,\ldots,N-1$.
These interior cardinal functions are distinct from the full-grid cardinal
functions $\ell_j(\theta)$. We define the corresponding Fej\'er-II
weights by
\begin{equation}
W_j=\int_0^\pi \lambda_j(\theta)\,d\theta,
\qquad j=1,\ldots,N-1 .
\end{equation}
The corresponding interior quadrature rule is
\begin{equation}
\int_0^\pi f(\theta)\,d\theta
\approx
\sum_{j=1}^{N-1}W_j f(\theta_j).
\end{equation}
The discretized delta function $\Delta_{ij}$ is defined by the rule that
\begin{equation}
f(\theta_i)=\int_0^\pi \delta(\theta_i-\theta_s) f(\theta_s) \, d\theta_s = \sum_{j=1}^{N-1} W_j \Delta_{ij} f(\theta_j).
\end{equation}
The second equality holds only in the large-$N$ limit, but imposing it at finite $N$ gives the condition $ \Delta_{ij} W_j=\delta_{ij}$ (no sum over $j$), or equivalently
\begin{align}
    \Delta = \textrm{diag}(W_1^{-1},\dots, W_{N-1}^{-1}).
\end{align}
As the number of collocation points $N \to \infty$, this matrix representation converges to the continuum delta function (see \cite{Blechta2024GreenOperator}).  We therefore define the discretized Green function $G$ to obey
\begin{equation}
L(k)\,G(r,r_s;k)
=
\Delta_n \,\delta(r-r_s),
\end{equation}
where we define the $n \times n$ matrix
\begin{align}
   \Delta_n= \begin{pmatrix}\Delta & 0 \\ 0 & \Delta \end{pmatrix}.
\end{align}

The continuum Green function $\mathcal{G}$ can be reconstructed using the interior and full-grid Cardinal functions.  We define the $ 2(N+1)\times 2$ evaluation matrix $E(\theta)$ by placing the full set of Chebyshev cardinal functions in each spin sector,
\begin{equation}
E(\theta) = 
\begin{pmatrix}
\ell_0(\theta) & 0 \\
\vdots & \vdots \\
\ell_N(\theta) & 0 \\
0 & \ell_0(\theta) \\
\vdots & \vdots \\
0 & \ell_N(\theta)
\end{pmatrix}.
\end{equation}
and similarly for the interior cardinal functions at the source point,
\begin{equation}
F(\theta_s) = 
\begin{pmatrix}
\lambda_1(\theta_s) & 0 \\
\vdots & \vdots \\
\lambda_{N-1}(\theta_s) & 0 \\
0 & \lambda_1(\theta_s) \\
\vdots & \vdots \\
0 & \lambda_{N-1}(\theta_s)
\end{pmatrix}.
\end{equation}
We may then write the continuum Green function interpolant as the matrix expression
\begin{equation}
(\mathcal{I}_N \mathcal{G})(r,\theta; r_s, \theta_s; k) = E(\theta)^T\mathcal{P} G(r, r_s; k) F(\theta_s).
\end{equation}

We construct the Green matrix $G$ from homogeneous solutions.  The interior angular collocation of a field $\Psi$ is the $n$-dimensional vector $u$ defined by
\begin{equation}\label{def:U}
u(r)=
\begin{pmatrix}
u_{-1}(r)\\
u_{-2}(r)
\end{pmatrix}
=
\begin{pmatrix}
\psi_{-1}(r,\theta_1)\\
\vdots\\
\psi_{-1}(r,\theta_{N-1})\\
\psi_{-2}(r,\theta_1)\\
\vdots \\
\psi_{-2}(r,\theta_{N-1})
\end{pmatrix}.
\end{equation}
Working in first-order form
\begin{equation}
Y(r)=
\begin{pmatrix}
u(r)\\
u'(r)
\end{pmatrix},
\end{equation}
the homogeneous solutions obey  (see \eqref{Lk})
\begin{equation}\label{linear-prop}
Y'(r)=\mathbb M(r,k)Y(r),
\end{equation}
with
\begin{equation}\label{def:M}
\mathbb M(r,k)=
\begin{pmatrix}
0 & I_n\\
-A_2^{-1}A_0 & -A_2^{-1}A_1
\end{pmatrix},
\end{equation}
where $I_n$ is the $n$ dimensional identity matrix. 

The source matching conditions follow by integrating the second-order
equation across $r=r_s$. Since the source contains $\delta(r-r_s)$ but no
derivative of a delta function, the field is continuous, while the first derivative has a jump,
\begin{equation}\label{eq:jump-cond}
[G]_{r_s}=0,
\qquad
A_2(r_s,k)[\partial_r G]_{r_s}=\Delta_n.
\end{equation}

The retarded boundary condition is imposed by choosing the appropriate
homogeneous subspaces on the two sides of the source.  Of the $2n$ homogeneous solutions, we expect $n$ that are purely ingoing at the horizon and $n$ that are purely outgoing at infinity.  From a set of $n$ linearly independent solutions $u_{(i)}(r)$ that are purely ingoing at the horizon we may form an $n\times n$ ``basis matrix''
\begin{equation}
U_H(r) = 
\begin{pmatrix} 
\vert & \vert & & \vert \\ 
{u}_{(1)}(r) & {u}_{(2)}(r) & \dots & {u}_{(n)}(r) \\ 
\vert & \vert & & \vert 
\end{pmatrix}.
\end{equation}
Similarly, we let $U_\infty(r)$ be an $n \times n$ basis matrix whose columns span the solutions that are purely outgoing at infinity.  For $r<r_s$, each column of the Green matrix is a homogeneous solution ingoing at the horizon.  We may thus write $G=U_H c_H$ in this region, where $c_H$ is an $n\times n$ coefficient matrix.  Similarly, for $r>r_s$ we have $G=U_\infty c_\infty$, i.e.
\begin{equation}
G(r,r_s;k)
=
\begin{cases}
U_H(r)c_H, & r<r_s,\\
U_\infty(r)c_\infty, & r>r_s.
\end{cases}
\end{equation}
The jump conditions \eqref{eq:jump-cond} then become
\begin{equation}
\begin{pmatrix}
U_H & -U_\infty\\
-A_2 U_H' & A_2 U_\infty'
\end{pmatrix}_{r=r_s}
\begin{pmatrix}
c_H\\
c_\infty
\end{pmatrix}
=
\begin{pmatrix}
0\\
\Delta_n
\end{pmatrix}.
\end{equation}
Once $U_H$ and $U_\infty$ are found, $c_H$ and $c_\infty$ are determined by solving this linear system.

The homogeneous-solution subspaces are determined from the eigenvalues of the matrix $\mathbb M(r,k)$ defined in Eq. (\ref{def:M}) \cite{AscherMattheijRussell1995BVP}.  The point is that a local eigenvector of the first-order system determines a local radial exponent.  If we write the eigenvalue equation at $r=r_0$ as
\begin{equation}
\mathbb M(r_0,k)
\begin{pmatrix}
v\\ d
\end{pmatrix}
=
\lambda
\begin{pmatrix}
v\\ d
\end{pmatrix},
\end{equation}
then the upper block implies $d=\lambda v$, i.e., $u'(r_0) = \lambda u(r_0)$.  At large $r$, the solutions behave as $u(r)\sim e^{\pm i \omega r}$.  A purely outgoing/ingoing solution is proportional to $e^{\pm i \omega r}$ and is a local eigenvector with eigenvalue $\lambda=\pm i \omega$, with the upper sign for outgoing and the lower sign for ingoing.  To isolate the outgoing sector, we choose a sufficiently large value of $r$, find the $n$ eigenvalues closest to $+i \omega$, and use the corresponding eigenvectors as initial data to evolve inward to the source point $r_s$.  The collection of solutions forms the matrix $U_\infty$.  The analogous strategy is employed near the horizon, where the solutions behave as $e^{\pm i k a /(2x)}$ with approximate eigenvalues $\lambda=\mp ik a/(2x^2)$.  To select the ingoing branch, we choose a sufficiently small value of $x$, find the $n$ eigenvalues closest to $\lambda=- ik a/(2x^2)$, and use the corresponding eigenvectors as initial data to evolve outward to the source point $r_s$. The collection of solutions forms the matrix $U_H$.  The actual values used in stated results are
\begin{equation}
x_{\rm in}=2\times10^{-9},\quad
r_s=2M,\quad
r_{\rm out}=30M.
\end{equation}

For numerical purposes, at each radius $r_j$ we represent each admissible $n$-dimensional solution
subspace by a $2n\times n$ matrix $Q_j$.  We use uniform steps in $\log x$. To prevent the fastest-growing solution from dominating the other basis vectors,
we orthonormalize after each radial step. If $\Phi_j$ is the
propagator from $r_j$ to $r_{j+1}$, we write
\begin{equation}
    \Phi_j Q_j
    = Q_{j+1}R_{j+1},
    \qquad
    Q_{j+1}^{\dagger}Q_{j+1}=I_n,
\end{equation}
where $R$ is upper-triangular.  Because $R_{j+1}$ is invertible, the QR decomposition changes only the
basis, not the solution subspace it spans.  

For the horizon-side solution, we also store the triangular factors
$R_j$. Suppose matching at $r_s=r_J$ gives coefficients
$c_{H,J}$ in the final basis $Q_J$. The coefficients in the
earlier QR bases are then obtained recursively from
\begin{equation}
    c_{H,j}=R_{j+1}^{-1}c_{H,j+1},
\end{equation}
and the Green matrix is reconstructed in the local basis at a point $r_j$ as
\begin{equation}
    G(r_j,r_s;k)=[Q_j]_u\,c_{H,j},
\end{equation}
where $[Q_j]_u$ denotes the upper $n$ rows of $Q_j$.  This allows us to reconstruct the Green matrix in the overlap regime of interest.

\subsection{Overlap region behavior}

By analogy with the Kerr spacetime result \eqref{transfer_overlap}, at sufficiently small $k$ we expect the Green matrix to expressible as a sum over 
(possibly complex) powers of $x$ in the overlap region $k \ll x \ll 1$.  To test this prediction we extract individual powers from the sampled Green matrix using a 
 block-Hankel dynamic mode decomposition \cite{Arbabi2016}, treating translation
in $t=\log x$ as the evolution variable with timestep $\Delta t$.  We use a uniformly spaced grid,
\begin{align}
    t_k=t_0+k\Delta t,
\end{align}
indexed by an integer $k$ (distinct from the near-corotation parameter used elsewhere in this paper).  

We are able to reliably extract powers from just $\psi_{-1}$, so we restrict to this sector for simplicity.  We vectorize the 
$\psi_{-1}$ block into a snapshot $Y_k\in\mathbb C^p$.  The full discretized Green matrix has $n$ field rows and $n$ source columns; the retained block has $n/2$ field rows and all $n$ source columns, giving $p=n^2/2$ entries per
snapshot.  

In the overlap region these snapshots are modeled as a finite sum over exponentials
\begin{equation}\label{Ykansatz}
  Y_k=\sum_{j=1}^{J}A_j e^{\lambda_j k \Delta t},
  \qquad A_j\in\mathbb C^p,
\end{equation}
where the number of terms $J$ is determined as part of the fitting process.  It is helpful to define
\begin{align}
    \mu_j=e^{\lambda_j \Delta t},
\end{align}
so that
\begin{align}\label{Ymu}
      Y_k=\sum_{j=1}^{J}A_j (\mu_j)^k.
\end{align}
For each $k$ we also form a vectorized interval containing the $d$ snapshots beginning with $Y_k$,
\begin{align}
     z_k=
 \begin{pmatrix}Y_k\\Y_{k+1}\\\vdots\\Y_{k+d-1}\end{pmatrix}
 \in\mathbb C^{pd}.
\end{align}
From the ansatz \eqref{Ymu} we have
\begin{align}\label{zbzbzb}
    z_k=\sum_{j=1}^J b_j(\mu_j)^k,
\end{align}
where the coefficients are rescaled from $A_j$ as
\begin{align}
      b_j=
 \begin{pmatrix}
 A_j\\ \mu_jA_j\\ \vdots\\ \mu_j^{d-1}A_j
 \end{pmatrix}.
\end{align}
Subsequent intervals $z_k$ and $z_{k+1}$ are both linear combinations of the same vectors $b_j$, and hence are related by a matrix $T$,
\begin{align}\label{Tisgreat}
    z_{k+1} = T z_k.
\end{align}
Then from \eqref{zbzbzb} we have
\begin{align}
    \sum_{j=1}^{J}(\mu_j)^k(b_j \mu_j - T b_j)=0.
\end{align}
We then assume that the $\mu_j$ are independent, so that 
\begin{align}\label{eq:full-shift-eigenpair}
    T b_j=\mu_j b_j.
\end{align}
Thus the evolution factors $\mu_j$ appear as eigenvalues of $T$.  The corresponding radial exponents are
\begin{equation}
  \lambda_j=\frac{\log\mu_j}{\Delta t}.
\end{equation}

The above data consists of a single interval with $d$ values of $t$ starting at $t_k$.  To incorporate all such intervals we define matrices whose columns are the intervals,
\begin{align}
 X&=\begin{pmatrix}z_0&\cdots&z_{K-d-1}\end{pmatrix},\\
 X'& =\begin{pmatrix}z_1&\cdots&z_{K-d}\end{pmatrix}.
\end{align}
The evolution \eqref{Tisgreat} acts on each column, so we have the full matrix equation
\begin{align}
    X'=TX.
\end{align}

These expressions hold exactly when the ansatz \eqref{Ykansatz} is exact.  When we instead form $X$ and $X'$ from the numerical data, we seek the matrix $T$ that minimizes the Frobenius norm of $X'-TX$, which may be expressed as $T=X'X^+$ in terms of the Moore-Penrose pseudoinverse $X^+$.  In practice, we work with rank-$r$ SVD descriptions of these matrices, treating $r$ as a parameter to be varied.

The number of exponents $J$ is not supplied to the fit.  We repeat the
extraction over delay depths $d=2,4,6,8$ and requested SVD ranks
$r=4,6,8,10,12,14$, giving 24 configurations.  In each fit, the requested
rank is capped at the numerical rank, defined as the number of singular
values satisfying $\sigma_i>10^{-12}\sigma_1$.  Therefore $r$ sets  the
maximum number of components that a single fit can resolve and the reported
mode count is inferred from the persistent clusters defined below.

For each $(d,r)$ fit, we diagonalize the rank reduced $\tilde{T}_r$ and retain the eigenvalues $\mu_j$ with corresponding $\lambda_j$ satisfying $-2<\operatorname{Re}\lambda_j<4$, 
$|\operatorname{Im}\lambda_j|<\pi/|\Delta t|$, and a normalized weight condition $W_j\geq0.03$,  where $W_j=|c_j|\,\lVert b_j\rVert/\max_\ell\bigl(|c_\ell|\,\lVert b_\ell\rVert\bigr)$ is defined in terms of the coefficients $c_j$ of a least-squares fit $z_0\simeq\sum_j c_j b_j$, where $b_j$ are the numerical eigenvectors.   The survivors from all 24 fits are grouped by
single-linkage clustering in the complex $\lambda$ plane, with Euclidean
distance threshold $0.09$.  We regard an exponent as resolved only if its
cluster is represented in at least 12 distinct $(d,r)$ configurations.  Its reported value is the strength-weighted cluster mean.

We also vary the size of the collocation grid, using $N=8$ and $N=14$.  We choose the value of $k$ to be numerically small,
\begin{align}
 k=2\times10^{-7}.
\end{align}
We then consider $x$ in the range
\begin{equation}
 8\times10^{-4}\leq x\leq1.9\times10^{-2},
\end{equation}
which satisfies the overlap-region condition $k \ll x \ll 1$.  We divide this region into 32 grid points with logarithmic spacing.

For each choice of $Q$ and $m$, we recover the exponents $\lambda$ and compare to the weights determined with the purely angular method of App.~\ref{app:numerical_exp}.  For the cases considered, we recover all modes with the leading power-law scaling.  That is, when there are principal modes, we recover these modes, and when there is no principal mode, we recover the leading supplementary mode.  We now present a couple of examples.

At $Q/M=0.50$, $m=1$, and $N=14$, we recover a single exponent
$\lambda=0.910713-0.000036i$. The independently computed value from the
angular method of App.~\ref{app:numerical_exp} is $p_{60}=0.911669$, differing from the extracted
exponent by $1.27\times10^{-3}$.

At $Q/M=0.65$, $m=2$, and $N=14$, we recover two exponents
\begin{align}
 \lambda_- & =0.501072-1.817795i,\\
 \lambda_+& =0.501358+1.820517i.
\end{align}
Their mean phase magnitude is $1.8192$, and their mean real
part is $0.501215$, differing from $1/2$ by
$1.215\times10^{-3}$. The corresponding weights computed with the more accurate angular method are 
\begin{align}
 p_{-,60}&=0.50000000-1.82402987i,\\
 p_{+,60}&=0.50000000+1.82402987i.
\end{align}
Thus the sampled Green matrix reproduces the expected radial powers. We use the angular method of App.~\ref{app:numerical_exp} for all reported values of $h$ (e.g., Tab.~\ref{tab:kn-physical-h-grid}).  The less accurate results of this appendix are used only to confirm the appearance of the expected powers in the overlap region of the physical retarded Green function.

\bibliographystyle{apsrev4-2}
\bibliography{References}

\end{document}